\documentclass[fleqn,usenatbib]{mnras}

\usepackage{ae,aecompl}

\usepackage{graphicx}	
\usepackage{amsmath}	
\usepackage{amssymb}

\title[Late time light curve of SN\,2011fe]{The late-time light curve of the type Ia supernova SN\,2011fe}

\author[G. Dimitriadis et al.]{
G. Dimitriadis$^{1}$\thanks{E-mail: G.Dimitriadis@soton.ac.uk},
M. Sullivan$^{1}$,
W. Kerzendorf$^{2}$,
A. J. Ruiter$^{3,4}$,
I. R. Seitenzahl$^{3,4}$,
\newauthor
S. Taubenberger$^{2,5}$,
G. B. Doran$^{6}$,
A. Gal-Yam$^{7}$,
R. R. Laher$^{8}$,
K. Maguire$^{9}$,
\newauthor
P. Nugent$^{10,11}$,
E. O. Ofek$^{7}$
and J. Surace$^{8}$
\\
$^{1}$Department of Physics and Astronomy, University of Southampton, Southampton, SO17 1BJ, UK\\
$^{2}$European Southern Observatory, Karl-Schwarzschild-Str. 2, D-85748 Garching, Germany\\
$^{3}$ARC Centre of Excellence for All-sky Astrophysics (CAASTRO), Australian National University, Canberra, ACT 2611, Australia\\
$^{4}$Research School of Astronomy and Astrophysics, Australian National University, Canberra, ACT 2611, Australia\\
$^{5}$Max-Planck-Institut fur Astrophysik, Karl-Schwarzschild-Str. 1, D-85741 Garching, Germany\\
$^{6}$Jet Propulsion Laboratory, California Institute of Technology, Pasadena, CA 91109, USA\\
$^{7}$Benoziyo Center for Astrophysics, Weizmann Institute of Science, 76100 Rehovot, Israel\\
$^{8}$Spitzer Science Center, California Institute of Technology, M/S 314-6, Pasadena, CA 91125, USA\\
$^{9}$Astrophysics Research Centre, School of Mathematics and Physics, Queens University Belfast, Belfast BT7 1NN, UK\\
$^{10}$Department of Astronomy, University of California, Berkeley, CA 94720-3411, USA\\
$^{11}$Lawrence Berkeley National Laboratory, 1 Cyclotron Road, MS 50B-4206, Berkeley, CA 94720, USA\\
}

\date{Accepted XXX. Received YYY; in original form ZZZ}

\pubyear{2016}

\begin{document}
\label{firstpage}
\pagerange{\pageref{firstpage}--\pageref{lastpage}}
\maketitle

\begin{abstract}
  We present late-time optical $R$-band imaging data from the Palomar
  Transient Factory (PTF) for the nearby type Ia supernova SN\,2011fe.
  The stacked PTF light curve provides densely sampled coverage down
  to $R\simeq22$\,mag over 200 to 620 days past explosion.  Combining
  with literature data, we estimate the pseudo-bolometric light curve
  for this event from 200 to 1600 days after explosion, and
  constrain the likely near-infrared contribution.  This light
  curve shows a smooth decline consistent with radioactive decay,
  except over $\sim$450 to $\sim$600 days where the light curve
  appears to decrease faster than expected based on the radioactive
  isotopes presumed to be present, before flattening at around 600
  days.  We model the 200--1600\,d pseudo-bolometric light curve with
  the luminosity generated by the radioactive decay chains of
  $^{56}$Ni, $^{57}$Ni and $^{55}$Co, and find it is not consistent
  with models that have full positron trapping and no infrared
  catastrophe (IRC); some additional energy escape other than
  optical/near-IR photons is required. However, the light curve is 
  consistent with models that allow for positron escape
  (reaching 75\% by day 500) and/or an IRC (with 85\% of
  the flux emerging in non-optical wavelengths by day 600).  The presence of the
  $^{57}$Ni decay chain is robustly detected, but the $^{55}$Co decay
  chain is not formally required, with an upper mass limit estimated
  at 0.014\,M$_{\sun}$.  The measurement of the $^{57}$Ni/$^{56}$Ni
  mass ratio is subject to significant systematic uncertainties, but
  all of our fits require a high ratio $>0.031$ ($>1.3$ in solar
  abundances).

\end{abstract}


\section{Introduction}
\label{sec:introduction}

\vspace{0cm}

Type Ia supernovae (SNe Ia) have assumed a major role in cosmology as
standardisable candles over the last 20 years
\citep{riess98,perlmutter98} and continue to attract considerable
interest as astrophysical phenomena in their own right.  Although
decades of intensive study have led to significant breakthroughs in our
understanding of these systems, many questions remain unanswered
\citep[see the review of][]{2011NatCo...2E.350H}.  Perhaps the most
important is the underlying physical mechanism that leads to the
explosion \citep{2000ARA&A..38..191H,2013FrPhy...8..116H}, including
the configuration of the progenitor system. There is a general
theoretical and observational consensus that SNe Ia are caused by the
explosive thermonuclear burning of the degenerate material
\citep{1960ApJ...132..565H} of a carbon-oxygen white dwarf
star \citep{11fe,bloom12} in a binary system, with the light curve
primarily sustained by the radioactive decay of $^{56}$Ni synthesised
in the explosion
\citep{1969ApJ...157..623C,1994ApJ...426L..89K,2014Natur.512..406C}.
The nature of the companion star to the white dwarf, however, remains
uncertain \citep[see review of][]{Maoz14}.

The two most widely-discussed classical scenarios for the progenitor
system are the single degenerate scenario (SD), where the binary
consists of the accreting white dwarf with a non-degenerate companion
star \citep{Whelan73}, and the double degenerate Scenario (DD), where
the explosion originates from the `merger' of two white dwarfs
\citep{Iben84,1984ApJ...277..355W}. There continues to be significant
debate in the literature as to the relative frequency of the two
channels.  A promising approach to distinguish the competing scenarios
is the study of SNe Ia at late times via extremely late-time
photometry \citep{ropke12}.

The shape of a SN Ia bolometric light curve is driven by the energetic
output of the radioactive decay, and the opacity of the (homologously)
expanding ejecta into which the decay energy is deposited. The
principle contributor for the first few hundred days after explosion
is $^{56}$Co, generated from the decay of the most abundant
radioactive synthesised nuclide, $^{56}$Ni. Most of the energy in the
decay of $^{56}$Co is emitted in the form of $\gamma$-rays and the
rest in charged leptons, such as positrons and electrons, both of which are thermalised in the expanding
ejecta, creating optical and near-infrared (near-IR) photons. As the
ejecta expand and the column density decreases as $t^{-2}$, the ejecta
become transparent to the high-energy $\gamma$-rays which increasingly
escape thermalisation. This leads to an observed light curve decline
that is faster than that expected from the radioactive decay, until,
after about 75--250 days \citep{Childress15}, the contribution of positrons/electrons becomes
dominant \citep[e.g.,][]{1979ApJ...230L..37A} and the bolometric light
curve is expected to settle on to a decline broadly matching that of
$^{56}$Co.  This assumes that the majority of charged leptons get trapped in
the ejecta; the extent to which these get trapped is uncertain
\citep[e.g.,][]{Cappellaro97,Milne01}, but models with full positron/electron
trapping are consistent with the late-time optical/near-IR
observations of several SNe Ia
\citep[e.g.,][]{Sollerman04,Stritzinger07,leloudas09,graur15}.

At even later epochs, other longer-lived decay chains may also
contribute, in particular the chains of $^{57}$Ni and $^{55}$Co
\citep{Seitenzahl09,ropke12}. Both chains have decay steps with
half-lives of several hundred days, longer than $^{56}$Co, leading to
a slowly declining energy input and thus a predicted flattening in very
late-time ($>$500--600 days) SN Ia light curves. The first evidence
for such nuclides has been claimed for the nearby SN Ia SN\,2012cg
\citep{graur15}, and a slow-down in the decline of the optical/near-IR
light curves has been observed in other SNe Ia after $\sim$600 days
\citep{Cappellaro97,Sollerman04,leloudas09}. Spectral modelling of the
late-time SN\,2011fe spectra also seems to require additional energy
input from $^{57}$Co \citep{Fransson15}.

In principle, an accurate determination of the mass of these other
longer-lived nuclides provides a promising diagnostic tool to
distinguish between different explosion mechanisms and, in turn,
different progenitor scenarios. Theoretical modelling shows that
models with burning at higher central white dwarf densities predict
higher $\rm{^{57}Ni}/\rm{^{56}Ni}$ mass ratios, since $\rm{^{57}Ni}$
is a neutron-rich isotope that is produced in greater abundance in neutron
rich environments, such as near-Chandrasekhar mass explosion models.
Lower central density models, such as white dwarf mergers, generally
predict lower ratios \citep[for example, see][]{ropke12,Seitenzahl13}.

A major complication is the accurate determination of the bolometric
light curve. A robust theoretical prediction is the `infrared
catastrophe' (IRC), a predicted thermal instability that occurs at
$\sim$500\,d after maximum light \citep{Axelrod80}. During the nebular
phase up until the onset of any IRC, cooling of the SN ejecta results
from optical and near-IR transitions, dominated by \ion{Fe}{ii} and
\ion{Fe}{iii}.  However, as the ejecta expand and the temperature
decreases, at $\sim$2000\,K the cooling is instead predicted to become
dominated by fine structure lines in the mid-IR with excitation
temperatures of $\sim$500\,K. This is predicted to result in rapid
cooling, and the thermal emission moving from the optical/near-IR to
the mid/far-IR; only $\sim$5\% of the deposited energy would emerge in
the optical/near-IR at 1000 days after maximum light \citep{Fransson15}.

This rapid decrease in luminosity in the optical/near-IR has never
been observed in SN Ia light curves
\citep[e.g.,][]{Sollerman04,leloudas09}, which is not consistent with
predictions from model light curve codes
\citep{1998ApJ...496..946K,1998ApJ...497..431K}. The reason is
unclear. One possibility is that at least part of the SN ejecta is
kept at temperatures above that at which the IRC occurs due to
clumping in the ejecta \citep{leloudas09}. Lower density regions cool
more slowly and may therefore remain above the limit for the IRC, thus
maintaining some flux output in the optical/near-IR. Another
possibility is that optical/near-IR luminosity can be maintained by a
redistribution of UV flux into the optical wavebands, due to non-local
scattering and fluorescence \citep{Fransson15}. Other possible
complications include `freeze-out' \citep{1993ApJ...408L..25F}, where
the assumption that radiated luminosity is equal to the deposition of
energy from radioactive decay, breaks down.

The number of SNe Ia with the high-quality late-time data needed to
make progress with these issues remains small: events must be both
very nearby and (ideally) suffer from only minimal absorption from
dust along the line of sight, keeping them visible in the optical
until very late times. Moreover, it is advantageous if the SN
  is located in the outskirts of its host galaxy, avoiding possible
  crowding from nearby stars and simplifying the photometry at late
  times. In this paper, we study the late-time light curve of
SN\,2011fe \citep{11fe}, a low-extinction and spectroscopically normal
SN Ia in the face-on spiral galaxy M101 located at a distance of
6.4\,Mpc \citep{Shappee11}. The event has a remarkably rich data set,
making it the most well-studied SN Ia to date \citep[see][for a review
of the early scientific results from studies of SN\,2011fe]{Kasen13},
with excellent spectroscopic
  \citep{11fe,2012ApJ...752L..26P,Pereira13,Mazzali14,Mazzali15,taubenberger15,graham15,Shappee13,Patat13,Foley13}
  and photometric
  \citep{McClelland13,Matheson12,Richmond12,Munari13,Tsvetkov13,Zhang16,Shappee16}
  coverage. Some late-time photometric studies of SN\,2011fe have
already been performed \citep{Kerzendorf14,Zhang16,Shappee16}, and
here we present new data on SN\,2011fe from late-time monitoring by
the Palomar Transient Factory (PTF), combined with other ground-based
and \textit{Hubble Space Telescope} \textit{(HST)} data up until
around 1600 days after peak brightness.

In Section~\ref{sec:data}, we briefly introduce SN\,2011fe and
describe the new measurements of its light curve from PTF and the data
reduction techniques employed.  In Section~\ref{sec:11fe}, we present
the late time PTF light curve and describe how we construct a
pseudo-bolometric light curve. We introduce our analysis framework in
Section~\ref{sec:analysis11fe}, including correcting for photons
emitted at near-IR wavelengths, and present models for the bolometric
light curve, including evidence for the $^{57}$Ni decay chain.  In
Section~\ref{sec:discussion}, we discuss the implications of our
study. Throughout this paper, we adopt the AB magnitude system and a
Hubble constant of
$H_{0}=70\,\text{km}\,\mathrm{s}^{-1}\,\mathrm{Mpc}^{-1}$.

 
\section{Data}
\label{sec:data}

Our first task is to compile a complete photometric and spectroscopic
data set on SN\,2011fe. In this section, we discuss SN\,2011fe itself, and then
the new observations presented in this paper, i.e., a new late-time
light curve from the PTF survey. We also describe the significant
amount of photometric and spectroscopic data that we use from the
literature.

\subsection{SN\,2011fe and the PTF light curve}
\label{sec:sn-2011fe}

SN\,2011fe was discovered by PTF at MJD 55797.2, at an apparent
magnitude of $g=17.3$\,mag. Fits to the early time data give an
inferred epoch for the emergence of the first photons
 of MJD $55796.687\pm0.014$ \citep{11fe}. The
SiFTO light curve fitter \citep{Conley08} measures a stretch of
$0.98\pm0.01$ and a $B-V$ colour at maximum light of
$-0.07\pm0.02$\,mag \citep{Mazzali14}, together with an epoch of
maximum light in the rest-frame $B$-band of $55814.30\pm0.06$ (2011 September 10.4 UT); 
all phases in this paper are given relative to this epoch. These numbers
are all indicative of a normal SN Ia.  The SALT2 light curve fitter
\citep{2007A&A...466...11G} gives similar results \citep{Pereira13}.
The Milky Way extinction was $E(B-V)_\mathrm{MW}=0.009$\,mag, and a
measurement of the host galaxy extinction along the line-of-sight to
the SN gives $E(B-V)_\mathrm{host}=0.014\pm0.002$\,mag
\citep{2013A&A...549A..62P}.

The PTF \citep{Law09,Rau09} was a rolling, wide-field transient survey, operating from
2009 to 2012, using the CFH12k camera \citep{Rahmer08} mounted on the Samuel Oschin
48-in telescope at the Palomar Observatory (P48).  The typical cadence
of the survey was from hours up to 5 days, with data mostly taken in a
Mould $R$-band filter $R_\mathrm{P48}$ interspersed with some $g$-band
($g_\mathrm{P48}$) observations. The limiting magnitude in good
conditions was around $R_\mathrm{P48}\simeq20.5$\,mag. Although
SN\,2011fe exploded during the PTF survey, it continued to be
monitored after 2012 by the iPTF experiment (the intermediate Palomar
Transient Factory). The iPTF coverage began on 2013 January 1, at a
phase of $\simeq480$\,d. All the PTF and a portion of the iPTF data we analyse
are publicly available through the Infrared Science Archive 
at IPAC\footnote{\url{http://irsa.ipac.caltech.edu/}}.

For the PTF light curve of SN\,2011fe, we employ a pipeline used
extensively in earlier PTF papers \citep[see, e.g.,][for particular
details]{2014ApJ...789..104O,firth15}. The pipeline is based around a
classical image-subtraction concept. A deep reference image is
constructed using data prior to the first detection of SN\,2011fe,
which is then photometrically aligned, astrometrically registered
and re-sampled to each image containing the SN light (the science
image).  The point spread function (PSF) of the resampled and science
images are determined from isolated point sources in the images, and
the image with the best seeing is degraded to match the PSF in the worst
seeing image, using a similar technique to that of
\citet{2008MNRAS.386L..77B}. The reference image is then subtracted
from the science image, producing a `difference' image containing only
objects that have changed their flux level between the reference and
science image epochs.

\begin{figure*}
\begin{center}  
        \includegraphics[width=0.9\textwidth]{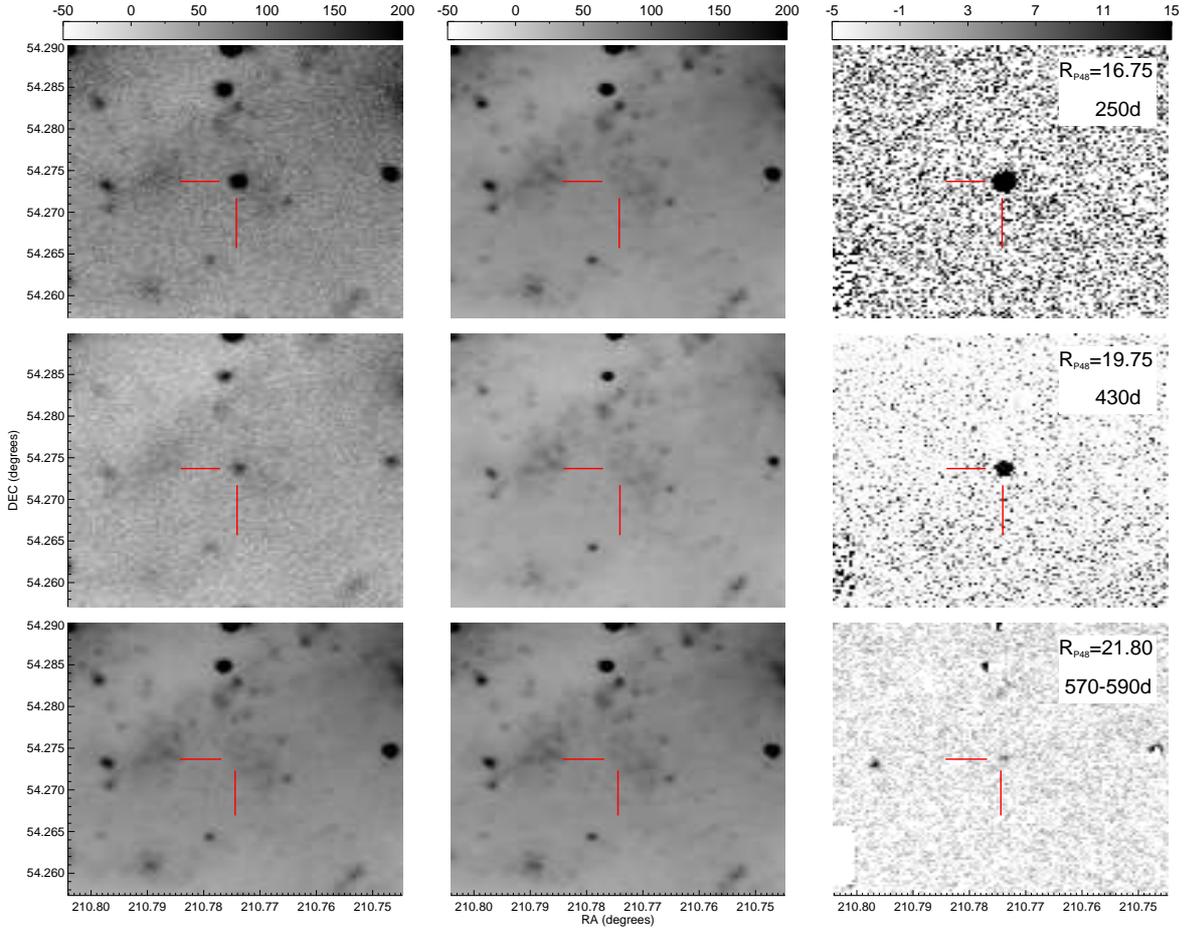}
        \caption{Late-time images of SN\,2011fe. The left column shows
          the science image, the centre column the reference image,
          and the right column the difference image. The corresponding
          phases of SN\,2011fe are (top to bottom) +250 days, +430
          days and an average of 40 images over 570-590 days, with the
          measured $R_\mathrm{P48}$ magnitudes indicated on the
          figure. The reference images indicate a low
          surface-brightness of M101 at the position of the SN
          \citep[see also the pre-explosion \textit{HST} imaging
          presented in][]{Li11}.}
\label{fig:11fe_images}
\end{center}  
\end{figure*}

The SN position is then measured from the difference images that have
a very high signal-to-noise (S/N) of the SN, and then PSF-fitting
photometry measured at this fixed position in all images after the SN
was first detected, calibrating to nearby stars from the Sloan Digital
Sky Survey \citep{2000AJ....120.1579Y} Data Release 12
\citep{2015ApJS..219...12A}. Note that SN\,2011fe was bright enough to
saturate the PTF camera when around peak brightness, and these
saturated images are discarded from our analysis. The bulk of the data
were taken in the $R_\mathrm{P48}$ filter (although $g_\mathrm{P48}$
data were also taken around maximum light), and thus we restrict the
PTF analysis to $R_\mathrm{P48}$ in this paper. Our PTF light curve of
SN\,2011fe is given in Table~\ref{tab:11fe_phot_log_ptf}, covering 819
$R_\mathrm{P48}$ images over PTF and iPTF, with typical science,
reference and difference images, shown in Fig.~\ref{fig:11fe_images}.

\subsection{Co-adding the PTF data}
\label{sec:co-adding-ptf}

SN\,2011fe is clearly detected in individual PTF images up until
around 400\,d after maximum light. After these epochs, the data
measured from the individual images, with typical 60\,s exposure
times, are mostly formal non-detections (i.e., $<3$\,$\sigma$
significance in any individual frame). However, much deeper limits can
be reached by co-adding the PTF data to increase the sensitivity. This
technique has previously been used on PTF data to successfully
detect pre-cursor events to core collapse SNe that lie below the
formal detection limit of PTF
\citep{2013Natur.494...65O,2014ApJ...789..104O,2014ApJ...782...42C,Strotjohann15}.

There are several choices as to how to perform this co-adding. The
individual science images could be co-added prior to image
subtraction; the individual difference images could be co-added prior
to the photometric measurement; or the co-adding can be performed on
the individual photometric data points produced by the pipeline. The
first two of these options have a significant disadvantage, in that
the PSF size and shape in the individual images may change
significantly over the averaging period, which may last up to several
tens of nights. Thus we instead choose to average the individual flux
measurements made from each PTF observation.

\begin{figure}
\begin{center}  
        \includegraphics[width=0.5\textwidth]{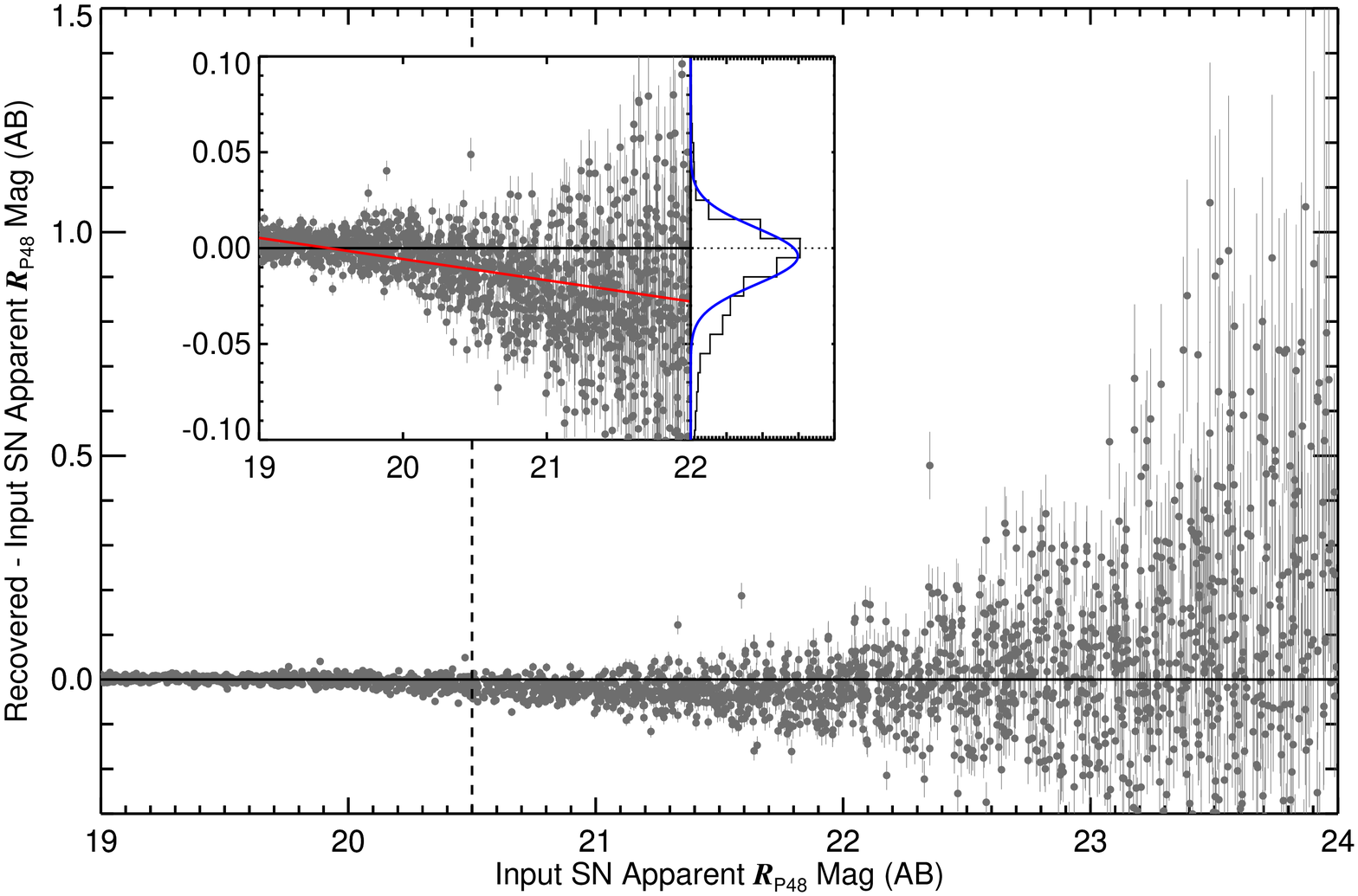}
        \caption{The fake SN tests on PTF imaging data (see Section
          \ref{sec:co-adding-ptf}). The main panel shows the
          difference between the recovered and input fake SN
          magnitudes for 2300 fake SNe, as a function of the input
          fake SN magnitude. The recovered magnitudes have been
          averaged over the period 200 to 700 days after maximum light
          of SN\,2011fe, with the vertical error-bars showing the
          uncertainty in the averaged magnitude. The vertical dashed
          line denotes the typical limiting magnitude in any
          individual PTF image ($R_\mathrm{P48}=20.5$). The inset
          shows the histogram of the magnitude differences over
          $19<R_\mathrm{P48}<22$, with the best-fitting Gaussian
          over-plotted on the histogram in blue. The red line denotes
          the linear correction we apply to the pipeline photometry.}
\label{fig:fakes}
\end{center}  
\end{figure}

In order to test the fidelity of the subtraction pipeline at these
faint flux levels, we insert artificial SNe (`fakes') into the
unsubtracted science images, and then test how accurately these fluxes
can be recovered after image subtraction and averaging of the PSF
photometric measurements.  We inserted 2300 fake point sources at
random positions, but with each fake having the same magnitude in each
frame chosen to lie over the range $19<R_\mathrm{P48}<24$.  We then
run the image subtraction pipeline on all images, and measure the PSF
flux of the fake SNe in each individual frame. These individual
measurements are then averaged together to give a final recovered
magnitude, exactly as for the real SN\,2011fe light curve. The result
of this test is shown in Fig.~\ref{fig:fakes}.

The pipeline generally performs well.  For the epoch range we are
particularly interested in ($+$200 to $+$700 days), when SN\,2011fe is expected
to be at $R_\mathrm{P48}\simeq19-22$, we calculate a mean offset of
$-0.006$ with a $0.019$ 1\,$\sigma$ standard deviation. The outliers on
the distribution are found mainly on the fainter side of the magnitude
space, especially at $\rm{mag}>21$, where naturally the pipeline
underperforms.

\subsection{Other data}
\label{sec:liter-data}

As well as the new PTF light curve data for SN\,2011fe, we also use
complementary photometric data from other sources. In particular, we
use data from \citet{Tsvetkov13} from the Crimean Astrophysical
Observatory out to $\sim$600\,d, and imaging from the Gemini-N
telescope from \citet{Kerzendorf14} at around $\sim$900\,d, for which
we make no attempt to correct for any background light and use it as published. We also
use photometric data from the Large Binocular Telescope (LBT) at $\sim$500-1600\,d and
 \textit{HST} imaging from the \citet{Shappee16} \textit{HST}
program at epochs $>$1000\,d (\textit{HST} proposals 13737 and 14166).

These data are compiled together in Table~\ref{tab:11fe_phot_log_ext}.
Note that, unlike the PTF and the LBT light curve, none of these data are based on
difference imaging, i.e. there is the possibility of contamination from
background light from M101.  However, deep pre-explosion \textit{HST}
imaging reveals no objects within 0.2" at the position of the SN to a 3-$\sigma$ limit of 26.6 mag, in F606W \citep{Li11}, and
thus the late-time \textit{HST} data will be only negligibly
contaminated by background light.

\begin{table*}{}
\caption{Literature $R$-band and near-IR photometric data for SN\,2011fe}
\label{tab:11fe_phot_log_ext}
\begin{tabular}{@{}lcccccl}
  \hline
  MJD&Phase &Telescope&Filter&Magnitude$^{b}$ &Magnitude&Reference\\
  & (days)$^{a}$ & & & & system &\\
  \hline
   
      $56049.50$&$234.93$&LBT&$J$&$17.18(0.10)$&Vega& \citet{Shappee16}\\
      $56257.50$&$442.76$&LBT&$J$&$17.42(0.04)$&Vega& \citet{Shappee16}\\
      $56332.50$&$517.70$&LBT&$J$&$18.23(0.18)$&Vega& \citet{Shappee16}\\

   $56337.48$&$522.67$&LBT&$R$&$22.14(0.02)$&Vega& \citet{Shappee16}\\
   $56419.46$&$604.59$&LBT&$R$&$21.74(0.03)$&Vega& \citet{Shappee16}\\

  $56440.81$&$625.92$&1m Crimean Astrophysical Observatory&$R$&$22.00(0.12)$&Vega&\citet{Tsvetkov13}\\
  $56441.84$&$626.95$&1m Crimean Astrophysical Observatory&$R$&$22.07(0.11)$&Vega&\citet{Tsvetkov13}\\
  
   $56449.21$&$634.32$&LBT&$R$&$21.91(0.02)$&Vega& \citet{Shappee16}\\
   $56453.26$&$638.36$&LBT&$R$&$21.95(0.02)$&Vega& \citet{Shappee16}\\

   $56637.53$&$822.48$&LBT&$R$&$23.19(0.04)$&Vega& \citet{Shappee16}\\

  $56743.00$&$927.87$&GMOS North&$r$&$24.01(0.14)$&AB&\citet{Kerzendorf14}\\
  
  $56812.35$&$997.17$&LBT&$R$&$24.31(0.08)$&Vega& \citet{Shappee16}\\
$56837.28$&$1022.08$&LBT&$R$&$24.54(0.08)$&Vega& \citet{Shappee16}\\
$56839.29$&$1024.08$&LBT&$R$&$24.27(0.06)$&Vega& \citet{Shappee16}\\
  
  $56939.00$&$1123.71$&\textit{HST}&F600LP&$24.84(0.02)$&Vega& \citet{Shappee16}\\
  $56939.00$&$1123.71$&\textit{HST}&F110W&$24.30(0.04)$&Vega& \citet{Shappee16}\\
  $56939.00$&$1123.71$&\textit{HST}&F160W&$22.37(0.02)$&Vega& \citet{Shappee16}\\

$57041.51$&$1226.14$&LBT&$R$&$25.34(0.20)$&Vega& \citet{Shappee16}\\
$57071.51$&$1256.11$&LBT&$R$&$25.46(0.14)$&Vega& \citet{Shappee16}\\

  $57118.00$&$1302.57$&\textit{HST}&F600LP&$25.55(0.05)$&Vega&\citet{Shappee16}\\
  $57118.00$&$1302.57$&\textit{HST}&F110W&$25.28(0.09)$&Vega& \citet{Shappee16}\\
  $57118.00$&$1302.57$&\textit{HST}&F160W&$23.12(0.02)$&Vega& \citet{Shappee16}\\

$57135.19$&$1319.75$&LBT&$R$&$25.51(0.18)$&Vega& \citet{Shappee16}\\
$57163.23$&$1347.76$&LBT&$R$&$26.07(0.26)$&Vega& \citet{Shappee16}\\

  $57217.00$&$1401.52$&\textit{HST}&F600LP&$25.82(0.04)$&Vega&\citet{Shappee16}\\
  $57217.00$&$1401.52$&\textit{HST}&F110W&$25.43(0.10)$&Vega& \citet{Shappee16}\\
  $57217.00$&$1401.52$&\textit{HST}&F160W&$23.41(0.04)$&Vega& \citet{Shappee16}\\

$57389.53$&$1573.88$&LBT&$R$&$26.39(0.41)$&Vega& \citet{Shappee16}\\
$57425.49$&$1609.81$&LBT&$R$&$26.24(0.42)$&Vega& \citet{Shappee16}\\

  $57437.50$&$1621.81$&\textit{HST}&F600LP&$26.51(0.09)$&Vega&\citet{Shappee16}\\
  $57437.50$&$1621.81$&\textit{HST}&F110W&$26.15(0.19)$&Vega& \citet{Shappee16}\\
  $57437.50$&$1621.81$&\textit{HST}&F160W&$24.09(0.09)$&Vega& \citet{Shappee16}\\

  \hline
  \multicolumn{7}{l}{$^a$Assuming maximum light at MJD 55814.30 \citep{Mazzali14}.  }\\
  \multicolumn{7}{l}{$^b$1-$\sigma$ uncertainties in parentheses.  }\\
 \end{tabular}
\end{table*}

Our analysis also requires spectroscopic data, for the calculation of
$s$-corrections, in order to correct the magnitudes estimated from
different telescopes and bandpasses to a common system
($R_\mathrm{P48}$), and pseudo-bolometric luminosities. A significant
amount of spectral data on SN\,2011fe at late-time already exist, and
we use optical data from the Lick Observatory 3-m telescope, the 4.2-m
William Herschel Telescope (WHT), and the 10-m Keck-I and Keck-II
telescopes, and near-IR data from LBT.  We take all spectra from the
WISeREP
archive\footnote{\url{http://wiserep.weizmann.ac.il/}}\citep{Yaron12},
and details are presented in Table~\ref{tab:11fe_spec_log}.

\begin{table*}{}
\caption{SN\,2011fe spectroscopy log}
\label{tab:11fe_spec_log}
\begin{tabular}{@{}lccccl}
\hline
   Date (UT)&Phase (days)$^a$&Telescope&Instrument&Wavelength Coverage (\AA)&Reference\\
\hline
  2012 04 02&$205$&Lick 3m&KAST&$3440-10268$&\citet{Mazzali15}\\
   2012 04 23&$226$&Lick 3m&KAST&$3438-10176$&\citet{Mazzali15}\\
   2012 04 27&$230$&Keck-I&LRIS&$3200-9991$&\citet{Mazzali15}\\   
   2012 05 01&$250$&LBT&Lucifer&$11700-13100$&\citet{Mazzali15}\\
   2012 05 01&$250$&LBT&Lucifer&$15500-17400$&\citet{Mazzali15}\\
   2012 05 01&$250$&LBT&Lucifer&$20500-23700$&\citet{Mazzali15}\\   
   2012 05 26&$259$&WHT&ISIS&$3498-9490$&\citet{Mazzali15}\\
   2012 06 25&$289$&WHT&ISIS&$3426-10268$&\citet{Mazzali15}\\
   2012 07 17&$311$&Lick 3m&KAST&$3458-10254$&\citet{Mazzali15}\\
   2012 08 23&$348$&Lick 3m&KAST&$3488-10238$&\citet{Mazzali15}\\
   2013 04 08&$576$&Keck-II&DEIMOS&$4450-9637$&\citet{graham15}\\
   2014 05 01&$964$&Keck-I&LRIS&$3077-10328$&\citet{graham15}\\
   2014 06 23&$1034$&LBT&MODS1&$3100-10495$&\citet{taubenberger15}\\
 \hline
 \multicolumn{6}{l}{$^a$Assuming maximum light at MJD 55814.30 \citep{Mazzali14}.  }\\
 \end{tabular}
\end{table*}


\section{The late-time light curve of SN 2011fe}
\label{sec:11fe}

\begin{figure*}
\begin{center}  
       \includegraphics[width=0.9\textwidth]{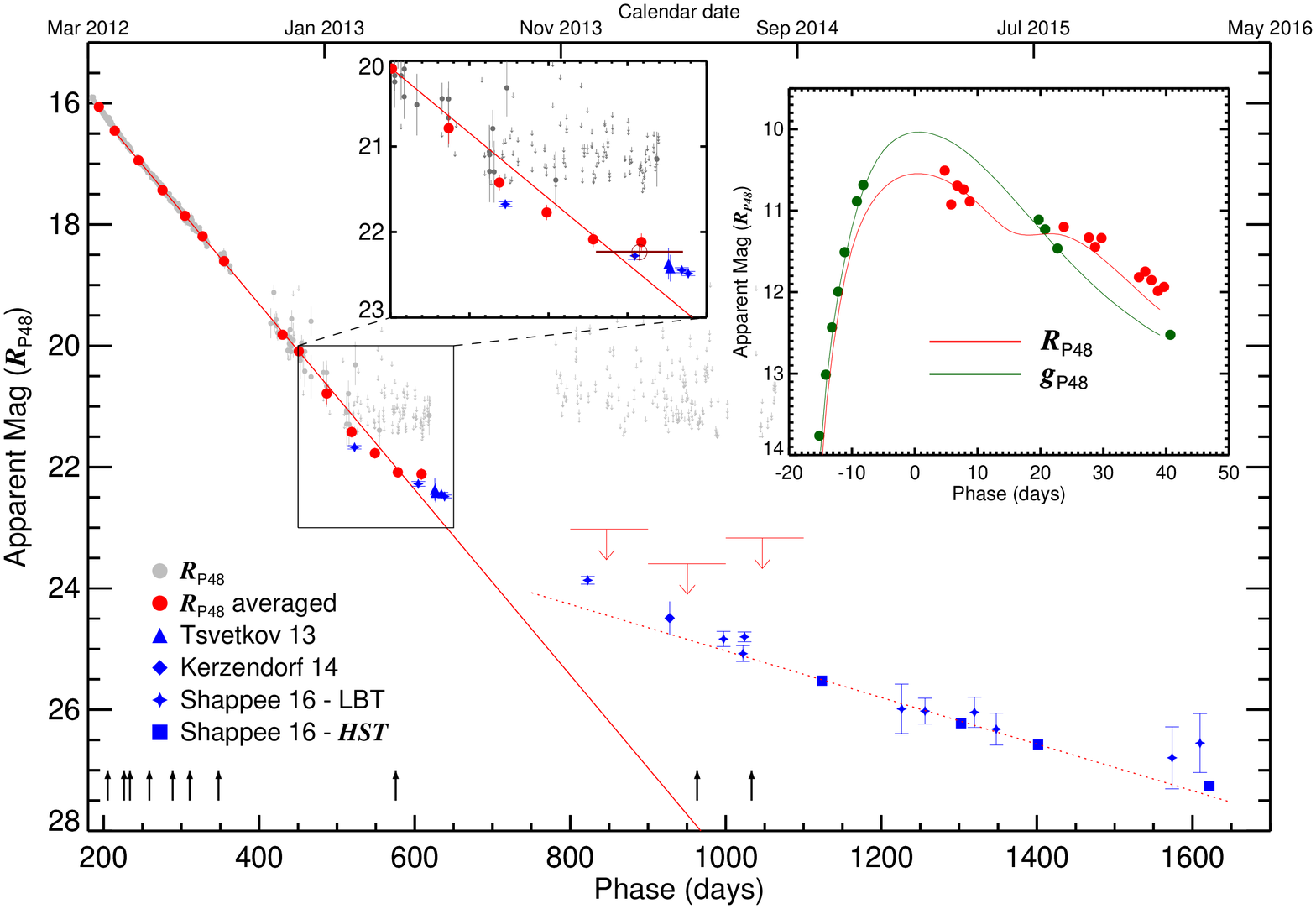}
       \caption{The SN\,2011fe late-time $R_\mathrm{P48}$-band light
         curve as a function of phase in days since maximum light. The
         light grey points show every individual PTF observation; when
         the significance of the detection is $<3$\,$\sigma$, an upper
         limit is plotted at 3\,$\sigma$. The red filled circles are
         averages of the PTF data in bins of 25 days. The literature
         photometric data are plotted as blue symbols and the sources
         are given in the legend; all are originally observed in some
         form of $R$-band filter (see
         Table~\ref{tab:11fe_phot_log_ext}) and $s$-corrected in the
         figure to $R_\mathrm{P48}$. The black vertical arrows
         correspond to the epochs of the spectral sample (see Table
         \ref{tab:11fe_spec_log}). The red solid line is a linear fit
         over the region of $+$200 to $+$600\,d, and the red dashed
         line a fit over $+$700 to $+$1620\,d. The left inset shows a
         zoom of the region between $+450$ to $+650$\,d, with the dark red
	open circle showing the averaged magnitude at the ~580-640d. The right
         inset shows the light curve around maximum light, together
         with fits from the SiFTO light curve fitter. Uncertainties
         are always plotted, but are often smaller than the data
         points.}
\label{fig:11fe_light}
\end{center}  
\end{figure*}

\begin{figure}
\begin{center}  
        \includegraphics[width=0.5\textwidth]{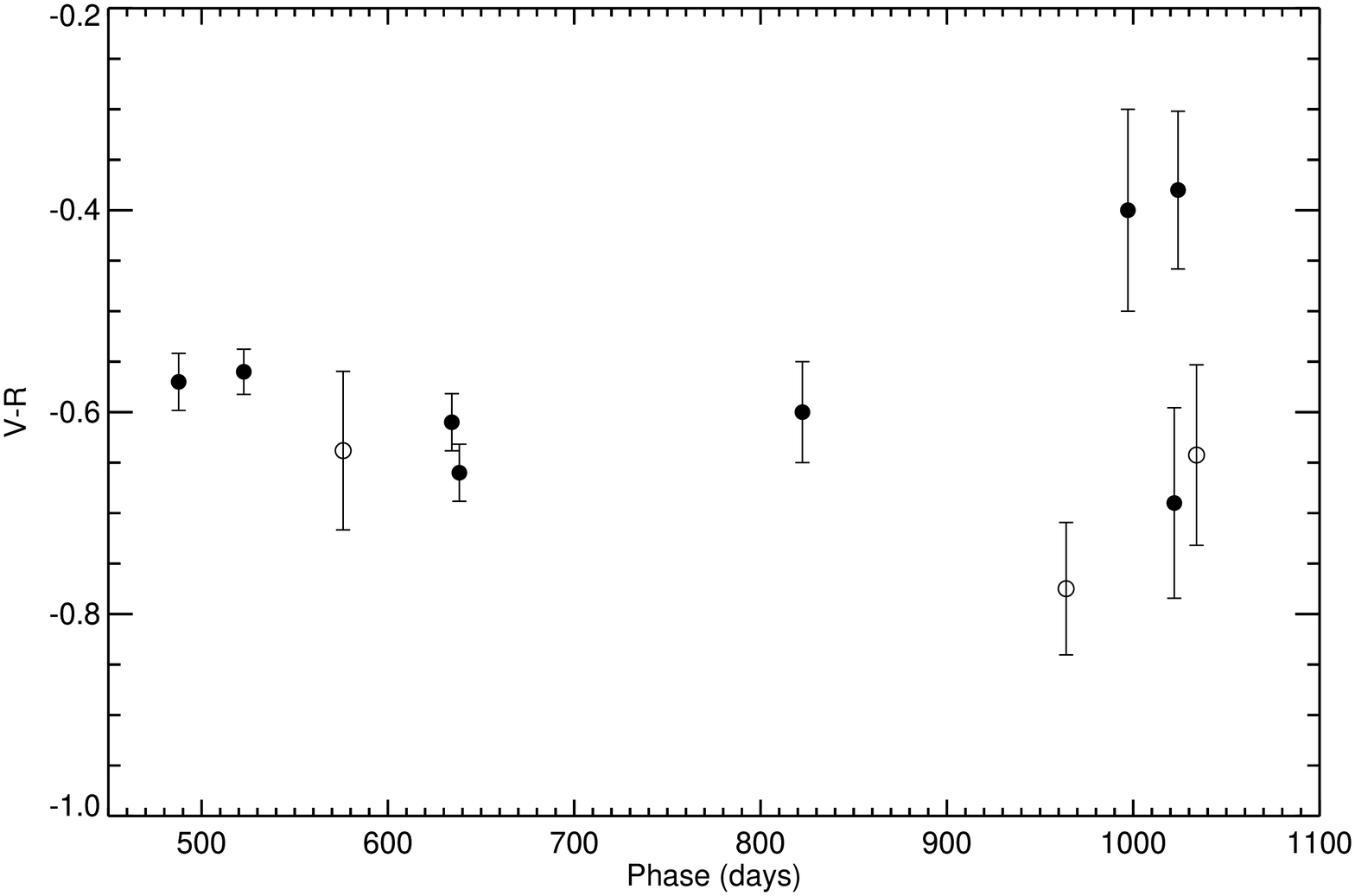}
        \caption{The $V$-$R$ colour of SN\,2011fe at late times. Full
          black circles are the photometric LBT colour from
          \citet{Shappee16}, and open black circles denoted the
          spectroscopic colour of the 576, 964 and 1034\,d spectra
          (see Table~\ref{tab:11fe_spec_log}).}

\label{fig:11fe_late_time_color}
\end{center}  
\end{figure}

In Fig.~\ref{fig:11fe_light}, we present the late-time photometry of
SN\,2011fe obtained from PTF. The PTF coverage is shown from around
$+200$\,d when the supernova was $R_\mathrm{P48}\sim16$, through to
$+1200$\,d, when the SN was not detected even in measurements averaged
over several weeks. There was no PTF coverage from around $+620$\,d to
$+780$\,d, as the field was not included in the iPTF survey during
this period. We also show the other ground- and space-based optical
data, where we have $s$-corrected to the $R_\mathrm{P48}$ filter using
the closest available spectrum. Such a procedure assumes that
  our spectra match the SEDs of the SN on the photometric epochs. We
  check this in Fig.~\ref{fig:11fe_late_time_color}, where we compare
  the late-time photometric $V$-$R$ colour of SN\,2011fe with the
  equivalent colour calculated from the available spectra. The
  photometric and `spectral' colours are consistent, with no
  indication of a mis-match between the colours calculated from the
  spectra and those from the photometry.

We identify three distinct regions of interest in the $R_\mathrm{P48}$
light curve. The first corresponds to 200 to 400\,d, the second
to 400 to 600\,d and the third to 600 to 1600\,d. During
the first and third regions, the light curve is declining linearly in
magnitude space (but with different slopes), qualitatively consistent with a light curve powered
by radioactive decay.  During the first period, we calculate a slope
of $1.53\pm0.09$ mag per 100 days \citep[consistent with similar
studies of SNe Ia in the $R$ band, see][]{Milne01,lair06} and in the
third region a slope of $0.37\pm0.12$ mag per 100 days. In the second
region, the light curve appears to deviate from a simple linear
radioactive decay, showing a faster decline for $\sim$100\,d followed
by an apparent flattening over the next $\sim$50\,d. The final PTF
measurement is consistent with the \citet{Tsvetkov13} $R$-band
measurement at 625\,d.

\subsection{A pseudo-bolometric light curve}
\label{sec:pseudo-bolomt-light}

A more detailed and quantitative investigation requires a bolometric
light curve covering wavelengths from the ultraviolet (UV) to the
infrared (IR). However, there are few published observations in the
near-IR of SN\,2011fe at these late phases: the latest spectroscopic
observation is at $\simeq$250\,d \citep{Mazzali15}, while for
  photometric observations, available data include LBT ($J$ band) and
  \textit{HST} imaging, both presented in \citet{Shappee16}. Thus a
true bolometric light curve is not possible to directly construct, and
we restrict ourselves to constructing a bolometric light curve based
only on the optical data where good spectral coverage does exist,
which we hereafter refer to as a `pseudo'-bolometric light curve. We
then correct for the likely near-IR contribution in our models, based
on data from SN\,2011fe and other events, when analysing this
pseudo-bolometric light curve. We describe the pseudo-bolometric light
curve in this section.

\begin{figure}
\begin{center}  
        \includegraphics[width=0.5\textwidth]{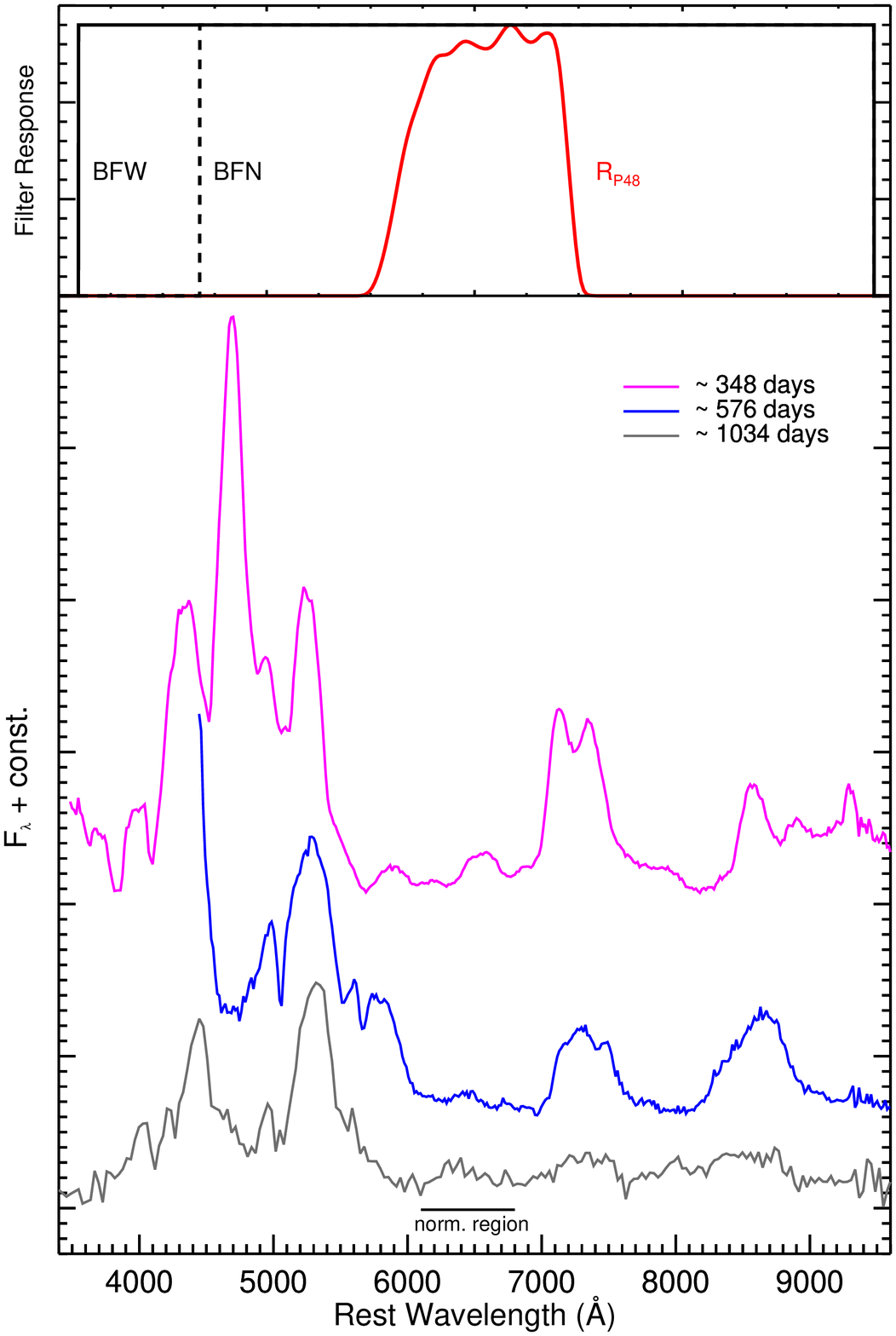}
        \caption{Lower panel: Comparison of three late-time optical
          spectra of SN\,2011fe at (from top to bottom) 348\,d,
          576\,d and 1034\,d. Upper panel: The filter responses of the
          $R_\mathrm{P48}$ filter (red), the wide bolometric box
          filter (BFW, solid black) and the narrow bolometric box
          filter (BFN, dashed black).}

\label{fig:11fe_spec}
\end{center}  
\end{figure}

Table~\ref{tab:11fe_spec_log} describes the published spectral data
available, and three of these late-time spectra, at $348$\,d, $576$\,d
and $1034$\,d, are shown in Fig.~\ref{fig:11fe_spec}. We introduce two
pseudo-bolometric box filters, a wide `BFW' filter from
3550-9500\,\AA, and a slightly narrower `BFN' filter covering
4500-9500\,\AA. The narrower BFN filter is designed for the narrower
wavelength coverage of the $576$\,d spectrum. These two filter
responses are also shown in Fig.~\ref{fig:11fe_spec}, together with
the $R_\mathrm{P48}$ filter response. The spectral features clearly
evolve over these phases, principally with the disappearance of the
[\ion{Fe}{iii}] line at 4700\,\AA. Most relevant for the
$R_\mathrm{P48}$ filter is the weakening of the 7200\,\AA\ feature
relative to the other stronger lines at bluer wavelengths
\citep[see][]{taubenberger15}.

We flux-calibrate the spectra in Table~\ref{tab:11fe_spec_log} by
estimating the $R_\mathrm{P48}$ magnitude on the epoch of the spectral
observation using linear interpolation. For the $+576$\,d spectrum
that coincides with the apparent break in the light curve, we average
the photometry from $+550$ to $+650$\,d instead of using the linear
interpolation. We rescale each spectrum so that it reproduces the
estimated $R_\mathrm{P48}$ apparent magnitude when integrated through
that filter, and then measure the pseudo-bolometric fluxes by
integrating through the BFW or BFN filters as appropriate. This is,
essentially, a flux-space `bolometric correction' for the
$R_\mathrm{P48}$ data determined on each epoch on which a spectrum
exists.

\begin{figure}
\begin{center}  
        \includegraphics[width=0.5\textwidth]{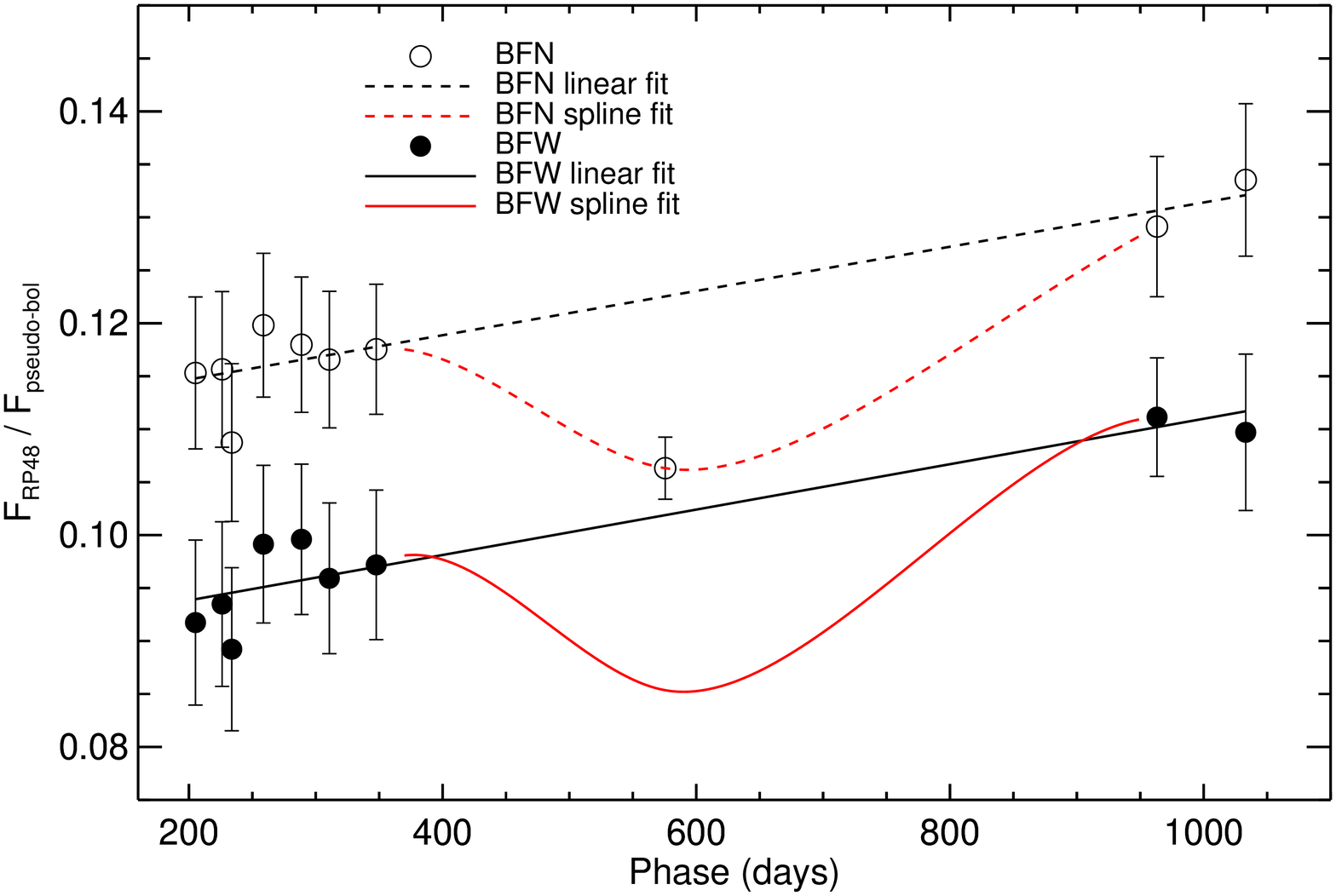}
        \caption{The ratio of the flux in the $R_\mathrm{P48}$ filter
          to that in the BFN filter (open circles) and the BFW filter
          (filled circles) for the spectral sequence in
          Table~\ref{tab:11fe_spec_log}.  Over-plotted are the linear
          (black) and spline (red) fits for the BFN filter (dashed
          lines) and the BFW filter (solid lines). The BFW spline is a
          rescaled version of the BFN fit, and is not independent.}
\label{fig:11fe_ratio}
\end{center}  
\end{figure}

We then extend the bolometric corrections made on epochs on which
spectra exist, to any epoch on which we have $R_\mathrm{P48}$ data. To
do this we check the stability of the bolometric correction by
plotting the ratio of the flux in $R_\mathrm{P48}$ to the flux in the
BFW and BFN filters in Fig.~\ref{fig:11fe_ratio}. The two ratios
evolve slowly as a function of time, with the relative flux in the
$R_\mathrm{P48}$ filter generally increasing. The exception to the
trend is the data point from the spectrum at $+576$\,d in the BFN
filter, where there is an apparent drop in the ratio. This is
temporally coincident with the apparent break in the
$R_\mathrm{P48}$-band light curve (Fig.~\ref{fig:11fe_light}).

We fit a simple smooth spline function to the BFN ratio data to
interpolate over this time period, superimposed on a linear fit to the
remaining epochs for interpolation at other epochs
(Fig.~\ref{fig:11fe_ratio}). We scale this spline fit to the BFW data,
and combined with a linear fit to the other BFW data. This process
allows us to translate the $R$-band data on any epoch into a
pseudo-bolometric luminosity, and thus construct a pseudo-bolometric
light curve. This light curve is shown in
Fig.~\ref{fig:11fe_bol_lightcurve_fully_trapped_no_irc} and presented in
Table~\ref{tab:11fe_lum_log}.

\begin{table}{}
\caption{The SN\,2011fe pseudo-bolometric luminosity light curve.}
\label{tab:11fe_lum_log}
\begin{center} 
\begin{tabular}{@{}lcl}
\hline
   Phase & $\log_{10}(L)$ &Source\\
   (days) & (erg\,s$^{-1}$)&(see Sec.~\ref{sec:pseudo-bolomt-light} for details)  \\
\hline
$205.00$&$40.676(0.035)$&Spec - \citet{Mazzali15}\\
$212.92$&$40.616(0.024)$&Phot - $R_\mathrm{P48}$\\
$226.00$&$40.532(0.033)$&Spec - \citet{Mazzali15}\\
$237.99$&$40.514(0.034)$&Phot - $R_\mathrm{P48}$\\
$259.00$&$40.312(0.037)$&Spec - \citet{Mazzali15}\\
$262.75$&$40.308(0.025)$&Phot - $R_\mathrm{P48}$\\
$287.20$&$40.151(0.026)$&Phot - $R_\mathrm{P48}$\\
$289.00$&$40.118(0.042)$&Spec - \citet{Mazzali15}\\
$311.00$&$40.002(0.039)$&Spec - \citet{Mazzali15}\\
$311.91$&$39.991(0.027)$&Phot - $R_\mathrm{P48}$\\
$331.12$&$39.872(0.027)$&Phot - $R_\mathrm{P48}$\\
$348.00$&$39.779(0.045)$&Spec - \citet{Mazzali15}\\
$354.90$&$39.732(0.028)$&Phot - $R_\mathrm{P48}$\\
$420.32$&$39.333(0.030)$&Phot - $R_\mathrm{P48}$\\
$440.69$&$39.218(0.031)$&Phot - $R_\mathrm{P48}$\\
$456.88$&$39.102(0.031)$&Phot - $R_\mathrm{P48}$\\
$486.84$&$38.892(0.032)$&Phot - $R_\mathrm{P48}$\\
$516.39$&$38.663(0.033)$&Phot - $R_\mathrm{P48}$\\
$522.67$&$38.546(0.034)$&Phot - \citet{Shappee16}\\
$536.90$&$38.548(0.034)$&Phot - $R_\mathrm{P48}$\\
$562.54$&$38.430(0.035)$&Phot - $R_\mathrm{P48}$\\
$576.00$&$38.399(0.035)$&Spec - \citet{graham15}\\
$585.51$&$38.450(0.036)$&Phot - $R_\mathrm{P48}$\\
$604.59$&$38.320(0.037)$&Phot - \citet{Shappee16}\\
$610.98$&$38.390(0.037)$&Phot - $R_\mathrm{P48}$\\
$625.92$&$38.278(0.037)$&Phot - \citet{Tsvetkov13}\\
$626.95$&$38.257(0.037)$&Phot - \citet{Tsvetkov13}\\
$634.32$&$38.247(0.038)$&Phot - \citet{Shappee16}\\
$638.36$&$38.230(0.038)$&Phot - \citet{Shappee16}\\
$822.48$&$37.604(0.045)$&Phot - \citet{Shappee16}\\
$927.87$&$37.382(0.050)$&Phot - \citet{Kerzendorf14}\\
$964.00$&$37.273(0.051)$&Spec - \citet{graham15}\\
$1034.00$&$37.056(0.058)$&Spec - \citet{taubenberger15}\\
$1123.71$&$36.897(0.066)$&Phot - \citet{Shappee16}-\textit{HST}\\
$1302.57$&$36.602(0.075)$&Phot - \citet{Shappee16}-\textit{HST}\\
$1401.52$&$36.453(0.087)$&Phot - \citet{Shappee16}-\textit{HST}\\
$1621.81$&$36.163(0.100)$&Phot - \citet{Shappee16}-\textit{HST}\\

 \hline
 \end{tabular}
 \end{center} 
\end{table}


\section{Analysis}
\label{sec:analysis11fe}

We now turn to the analysis of the radioactive decay chains that may
explain the shape and evolution of the pseudo-bolometric light curve
of SN\,2011fe. Our goals in this section are two-fold. The first goal
is to confirm that the shape of the pseudo-bolometric light curve can
be explained in terms of the radioactive decay chains that are
expected to be present, including $^{57}$Co and $^{55}$Fe 
\citep[e.g.,][]{Seitenzahl14,graur15,Shappee16}. The second goal is to attempt
to robustly determine the relative amounts of $^{56}$Co, $^{57}$Co and
$^{55}$Fe that are required to reproduce the pseudo-bolometric light
curve evolution.  We first introduce the framework of our modelling
and the assumptions, then discuss our procedures for adjusting the
model luminosity to account for near-IR contributions and finally fit
the pseudo-bolometric light curve.

\subsection{Analysis framework}
\label{sec:analysis-frame-work}

Several different decay chains may contribute at different levels
during the time period that we are sampling. These produce not only
$\gamma$-rays and positrons but also electrons and X-rays that can be
thermalised and deposit their energy in the expanding SN ejecta. At
the late-time epochs that we study here, when the SN ejecta become
increasingly optically thin, the delay between the energy deposition
from the radioactive decay and the emission of optical radiation
becomes short, and the light curve time-scale becomes driven by the
various decay chains.

We consider three decay chains that are likely to contribute to the
bolometric output:
\begin{align}
&^{56}\rm{Ni}\xrightarrow{t_{1/2}=6.08\:\rm{d}} {^{56}\rm{Co}}\xrightarrow{t_{1/2}=77.2\:\rm{d}}{^{56}\rm{Fe}}
\label{eq:ni56}
\end{align}
\begin{align}
&^{57}\rm{Ni}\xrightarrow{t_{1/2}=35.6\:\rm{h}} {^{57}\rm{Co}}\xrightarrow{t_{1/2}=271.2\:\rm{d}}{^{57}\rm{Fe}}
\label{eq:co57}
\end{align}
\begin{align}
&^{55}\rm{Co}\xrightarrow{t_{1/2}=17.53\:\rm{h}} {^{55}\rm{Fe}}\xrightarrow{t_{1/2}=999.67\:\rm{d}}{^{55}\rm{Mn}}
\label{eq:co55}
\end{align}
where the half-life is given in each case. Note that in each of these
chains, one of the decay steps is significantly longer than the other;
thus we make the approximation that only $^{56}$Co, $^{57}$Co and
$^{55}$Fe are important in our late-time ($t>200$\,d) light curve
analysis \citep[we neglect other decay chains such as $^{44}$Ti as
they will be sub-dominant for many years; e.g.,][]{Seitenzahl09}.

Under these approximations, the bolometric luminosity $L$ for a given
decay chain is

\begin{multline}
  L_{A}(t)=\frac{2.221}{A}\frac{\lambda_{A}}{\mathrm{days^{-1}}}\frac{M(A)}{M_{\sun}}\frac{q_{A}^{l}f_{A}^{l}(t)+q_{A}^{\gamma}f_{A}^{\gamma}(t)+q_{A}^{x}}{\rm{keV}} \\ \times\exp(-\lambda_{A} t)\times10^{43}\mathrm{erg\:s^{-1}}, 
  \label{eq:decaychainlum}
\end{multline}

where $t$ is the time since explosion, $A$ is the atomic number of the
decay chain, $M(A)$ is the initial mass synthesised, $\lambda_A$ is
the decay constant, $q_{A}^{l}$, $q_{A}^{\gamma}$ and $q_{A}^{x}$ are the average
energies per decay carried by the (electrically charged) leptons, $\gamma$-rays and X-rays, and
$f_{A}^{l}$ and $f_{A}^{\gamma}$ are the fractions of the leptons and
$\gamma$-rays absorbed in the ejecta, i.e. that contribute to the
luminosity. Under the assumption of homologous expansion, these
fractions are time-dependent and are given by
\begin{equation}
f_{A}^{l,\gamma}(t)=1-\exp\left[-\left(\frac{t_{A}^{l,\gamma}}{t}\right)^{2}\right],
\label{eq:trappingfraction}
\end{equation}
where $t_{A}^{l,\gamma}$ corresponds to the time when the optical
depth, $\tau_A^{l,\gamma}$, for positrons/leptons or $\gamma$-rays becomes
unity, i.e.  $\tau_A^{l,\gamma}=\left(t_A^{l,\gamma}/t\right)^2$. In
this framework, the effective opacity for each species,
$k_A^{l,\gamma}$, is proportional to $t_A^2$.  Under the leptonic
contribution channel, we include positrons, Auger and
internal conversion electrons.

The values for the decay energies and constants are presented in Table~\ref{tab:Decay_energies},
 and can be found at the National Nuclear Data
Center\footnote{\url{http://www.nndc.bnl.gov}}.

\begin{table}
\caption{Radioactive decay constants and energies per decay.}
\label{tab:Decay_energies}
\begin{center} 
\begin{tabular}{@{}lcccc}
  \hline
   Nucleus & $\lambda_{A}$ & $q^{\gamma}$ & $q^{x}$ & $q^{l}$  \\
       & $\mathrm{days^{-1}}$ & [keV] & [keV] & [keV]$^{a}$  \\
  \hline
  $\mathrm{^{56}Co}$& 8.975e-3 & 3606.0 & 1.587 & 124.61  \\
  $\mathrm{^{57}Co}$& 2.551e-3 & 121.6 & 3.6 & 17.814 \\
  $\mathrm{^{55}Fe}$& 6.916e-4 & -- & 1.635 & 3.973 \\
  \hline
  \multicolumn{2}{l}{$^a$Including $e^{+}$, Auger $e^{-}$ and IC $e^{-}$}\\
 
  \end{tabular}
 \end{center} 
\end{table}

In our analysis, we make the following simplifications:
\begin{enumerate}
\item We fix $t_{56}^{\gamma}=35$\,d. This parameter is not
  well-measured by our late-time pseudo-bolometric light curve, but is
  well-constrained in the literature: \citet{Sollerman04},
  \citet{Stritzinger07} and \citet{leloudas09} have calculated
  $t_{56}^{\gamma}$ for SN\,2000cx, SN\,2001el and SN\,2003hv to be
  $t_{56}^{\gamma}=31.5$, 35, and 32.5\,d respectively, while
  \citet{Zhang16} infer $t_{56}^{\gamma}=34.5$\,d for SN\,2011fe. We
  include the (small) contribution to the heating from $^{56}$Co
  $\gamma$-rays even at very late times.
 \item We fix $t_{57}^{\gamma}=160$\,d. This parameter is, again, not
  well-measured. The main $\gamma$-ray
 contribution from $^{57}$Co are the low-energy 14.4, 122.1
 and 136.5 keV lines. For the 14.4 keV, we assume complete trapping.
 For the 122.1 and 136.5 keV ones, we use a value for $t_{57}^{\gamma}$
 that suggests a 10\% trapping at 500 days.
\item We treat $t_{56}^{l}$, $t_{57}^{l}$ and $t_{55}^{l}$ as free
  parameters in some of the fits, i.e., we do not assume full trapping
  of the leptonic channel.
\item We assume complete trapping of the low energy X-rays.
\end{enumerate}

\subsection{Non-optical contribution}
\label{sec:non-optic-contr}

Our pseudo-bolometric light curve (Section
\ref{sec:pseudo-bolomt-light}) is based on optical data only. We must
therefore account for the radiation emerging at non-optical
wavelengths by either correcting the data, or by adjusting the model.
Here we choose to adjust the model, keeping our data as close to the
observations as possible.

\begin{figure}
\begin{center}  
        \includegraphics[width=0.5\textwidth]{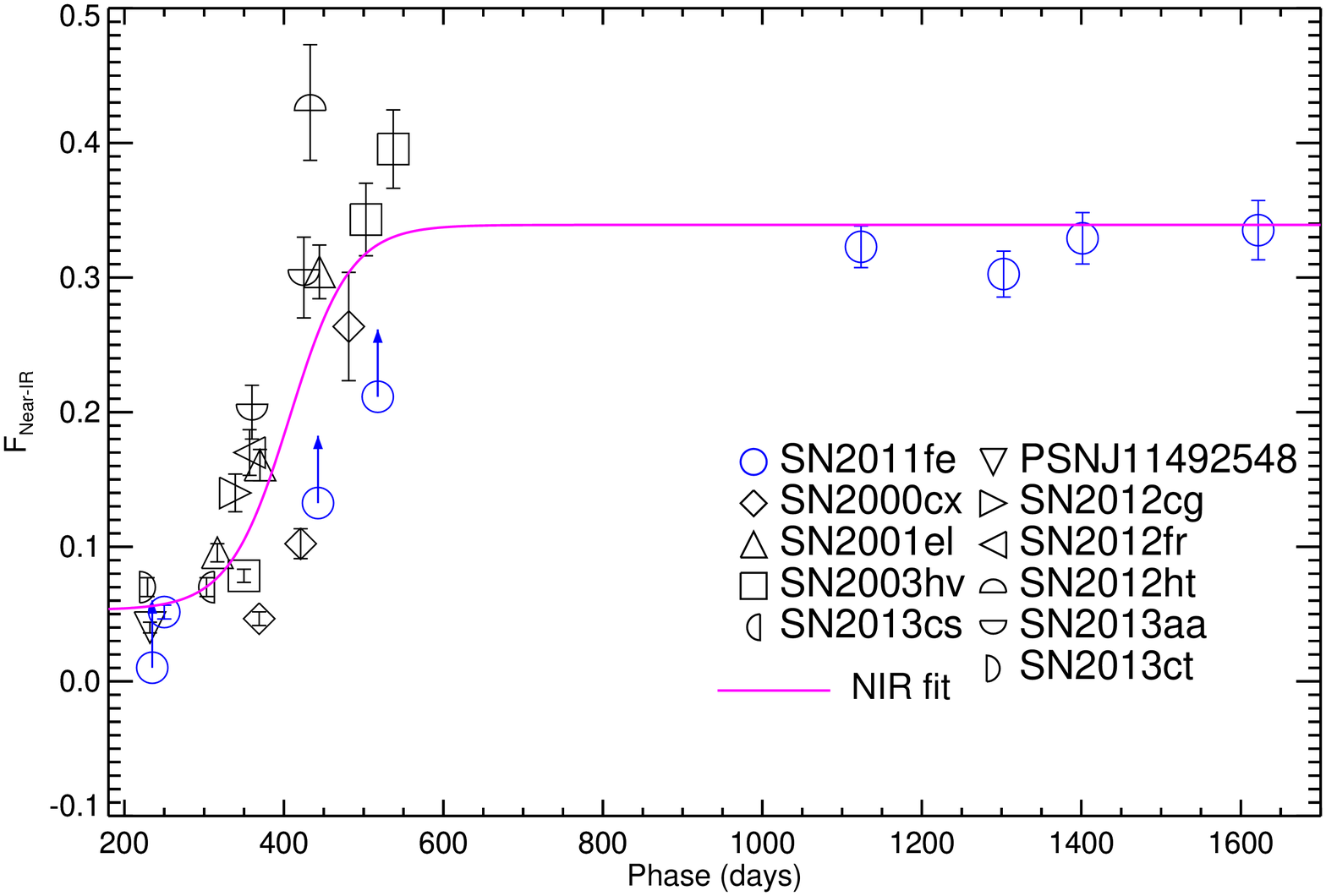}
        \caption{The Near-IR contribution, $F_\mathrm{near-IR}(t)$
          (equation~\ref{eq:NIR_contr}), for SNe Ia taken from the
          literature; see Section~\ref{sec:non-optic-contr}. The blue
          points show data for SN\,2011fe. The magenta line
          corresponds to the functional form that we adopt in our
          fitting to SN\,2011fe.}
\label{fig:noc_compare}
\end{center}  
\end{figure} 

In Fig.~\ref{fig:noc_compare}, we present the total near-IR
contribution for SN\,2000cx, SN\,2001el and SN\,2003hv based on
broad-band imaging \citep{Sollerman04,Stritzinger07,leloudas09}. We
also use near-IR estimates based on nebular spectra of 7 SNe Ia taken
from Maguire et al. (in preparation) and observed in the optical and
near-IR with the Very Large Telescope XShooter spectrograph. Finally,
we include a point for SN\,2011fe obtained by integrating the
available near-IR spectrum at 250\,d (Table~\ref{tab:11fe_spec_log}), 
the near-IR LBT data (plotted with upward arrows, since they are only
$J$-band observations) and the \textit{HST} data from \citet{Shappee16} at later
phases. Note for the literature SNe Ia the rapidly increasing near-IR
flux contribution over 400--600\,d, as flux is presumably
redistributed from the optical to the near-IR. This generally leads to
optical light curves declining more rapidly than near-IR light curves
over these epochs due to the decreasing temperature in the ejecta and
the consequently increased importance of the near-IR [\ion{Fe}{ii}]
lines \citep[e.g.][]{Sollerman04,leloudas09}.

The calculation of the near-IR contributions of SN\,2000cx, SN\,2001el
and SN\,2003hv were performed as follows. The published light-curve
magnitudes were extinction corrected and converted to fluxes, with
linear interpolation for the missing epochs. The optical
  fluxes were used to adjust the available spectra of SN\,2011fe so
  that the spectra reproduce the optical colours, using the `mangling'
  technique of \citet{Hsiao07} and \citet{Conley08}, and then
  integrated through the `BFW' filter (or the `BFN' filter, adding the
  appropriate correction from Section~\ref{sec:pseudo-bolomt-light}),
  thus calculating the optical part of the bolometric light curve. The
  near-IR $JHK$ fluxes were integrated over their effective filter
  area.  The near-IR contribution is then defined as:
  \begin{equation}
	F_\mathrm{nir}(t)=\frac{L_\mathrm{nir}(t)}{L_\mathrm{opt}(t)+L_\mathrm{nir}(t)}
\label{eq:NIR_contr}
\end{equation}

While the absolute contributions differ from SN to SN, the general
shape is similar, indicating a colour evolution over $\sim$300--600\,d
that shifts flux from the optical to near-IR wavelengths. We fit a
simple functional form, based on the logistic or sigmoid function, to
these data to model this near-IR contribution as a function of
time.While a more complex near-IR colour evolution is possible,
leading to a local maximum at $\simeq$650\,d, seen in SN\,2003hv
\citep[Fig. 8 of][]{leloudas09}, we will retain this simple functional
form, mainly because the last data point of SN\,2003hv is based on
extrapolating the light curve, under the assumption that the J-H and
H-K colours do not change, making the final calculation highly
speculative.

The final model luminosity, $L_\mathrm{model}(t)$, that we fit to our
pseudo-bolometric light curve is then
\begin{equation}
L_\mathrm{model}(t)=\left[L_{55}(t)+L_{56}(t)+L_{57}(t)\right]\times\left[1-F_\mathrm{nir}(t)\right]
\label{eq:finalmodel}
\end{equation}
where $L_{55}$, $L_{56}$ and $L_{57}$ are given by
equation~(\ref{eq:decaychainlum}), and $F_\mathrm{nir}$ is determined
as detailed above.

We also estimate the contribution of mid-IR emission over 500--600\,d
from the \textit{Spitzer} light curve obtained with the Infrared Array
Camera (IRAC) channels 1 and 2 (at 3.6 and 4.5\,$\micron$
respectively) presented in \citet{Johansson14}. We can only estimate
an upper limit, but find a total mid-IR contribution of $\sim$1 per
cent, indicating that little flux is emitted in the mid-IR region at
these epochs.  We defer a discussion of the far-IR region (and a
possible IRC) to the next section. Finally, we note that the
contribution at UV wavelengths smaller than 3500\,\AA\ is expected to
be small, since it is dominated by optically thick lines, and the only SN\,2011fe 
nebular UV spectrum shows little UV
flux (Friesen et al., in prep).

\subsection{Analysis of the bolometric light curve}
\label{sec:bol11fe}

\begin{figure*}
\begin{center}  
        \includegraphics[width=0.9\textwidth]{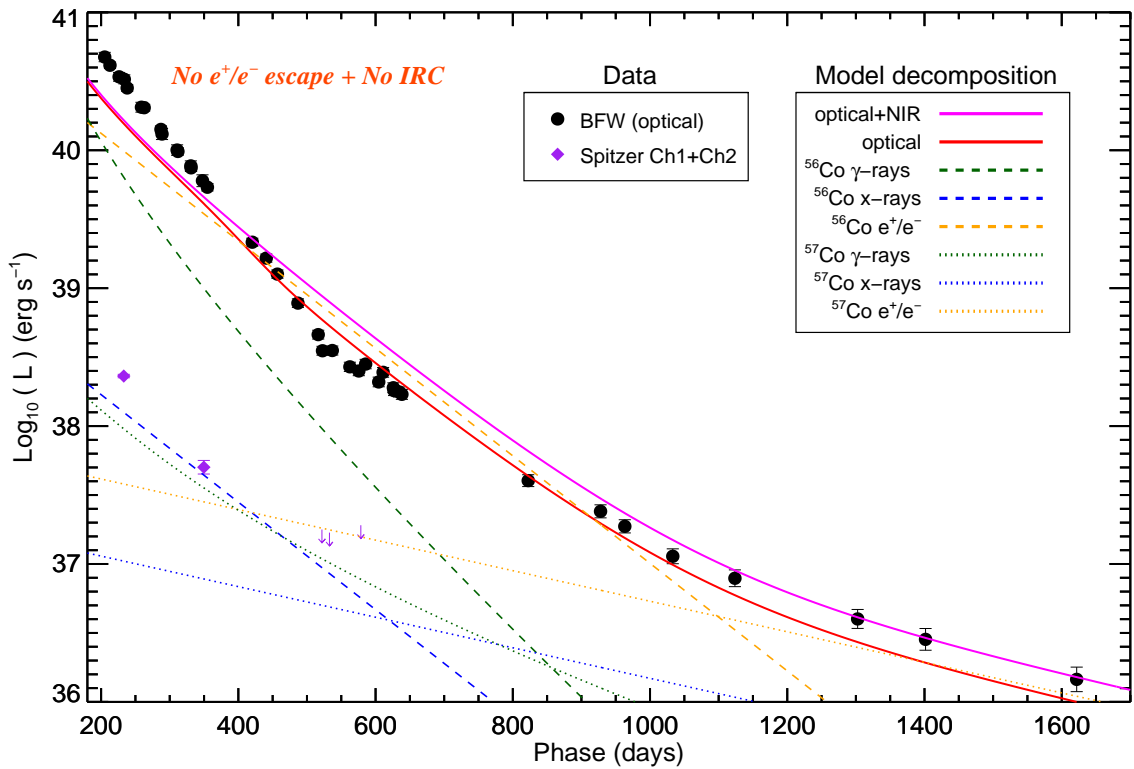}
        \caption{The pseudo-bolometric light curve of SN\,2011fe shown
          as filled black circles; see
          Section~\ref{sec:pseudo-bolomt-light} for details of the
          light curve construction. This figure shows the fit of the
          `Case 0' model (Section~\ref{sec:bol11fe}), assuming
          complete positron/electron trapping and no emergent far-IR
          flux. The magenta line is the total (optical plus near-IR)
          bolometric luminosity, and the red line is the final model
          fit to the pseudo-bolometric light curve data based on
          equation~(\ref{eq:finalmodel}), once the near-IR
          contribution is accounted for.  The mid-IR \textit{Spitzer}
          light curve \citep{Johansson14} is plotted as purple
          diamonds and upper limits. We include the decomposition of
          our model into the two decay chains of $A=56$ and $A=57$,
          separated into the $\gamma$-ray, X-ray and positron/electron
          components, shown in the legend.}
\label{fig:11fe_bol_lightcurve_fully_trapped_no_irc}
\end{center}  
\end{figure*}

The late-time pseudo-bolometric light curve of SN\,2011fe derived in
Section~\ref{sec:pseudo-bolomt-light} is shown in
Fig.~\ref{fig:11fe_bol_lightcurve_fully_trapped_no_irc}. In this section, we
consider various model fits to this light curve and attempt to
constrain the various fit parameters. 

We begin by fitting equation~\ref{eq:finalmodel}, including all
three decay chains (equation~\ref{eq:ni56} to equation~\ref{eq:co55}),
directly to our data, assuming full trapping of positrons/electrons and no flux
emerging at $\lambda>2.5$\,\micron\
(Fig.~\ref{fig:11fe_bol_lightcurve_fully_trapped_no_irc}), which we
refer to as \textit{Case 0}. The best-fitting parameters are reported
in Table~\ref{tab:Bol_fit_results}. This provides a very poor fit to
the data with a reduced $\chi^2$ (the $\chi^2$ per degree of freedom)
of $\chi^2_\mathrm{DOF}\simeq24$ for 36 degrees of freedom. The model cannot simultaneously fit
the data over 200--400\,d and over 900--1600\,d, and prefers a low
$^{56}$Ni mass ($M(56)=0.18$\,$M_{\sun}$).

\begin{table*}
\caption{The best-fitting parameters for the modelling of the pseudo-bolometric light curve.}
\label{tab:Bol_fit_results}
\begin{center} 
\begin{tabular}{@{}lccccccccc}
  \hline
  &  $M(56)$ & $M(57)$ & $M(55)$ & $M(57)/M(56)$ & $M(57)/M(56)$ & $t_{56}^{l}$ & $t_{57}^{l}$&$\chi^{2}_\mathrm{DOF}$&DOF\\
  Model  &  $(\mathrm{M_{\sun}})$ & $(\mathrm{M_{\sun}})$& $(\mathrm{M_{\sun}})$& & $(^{57}\mathrm{Fe}/^{56}\mathrm{Fe})_{\sun}$$^{a}$  & $(\mathrm{days})$& $(\mathrm{days})$&\\
  \hline
  Case 0 &$0.179(0.003)$&$0.004(0.001)$&$0.0(0.0)$&$0.022(0.003)$&$0.963(0.114)$&---&---&$24.09$&36\\
  
  Case 1 &$0.461(0.041)$&$0.014(0.005)$&$0.0(0.0)$&$0.031(0.011)$&$1.322(0.473)$&$233(30)$&$886(427)$&$2.47$&34\\
  Case 1, excluding 550-650d  &$0.442(0.035)$&$0.015(0.007)$&$0.0(0.0)$&$0.023(0.016)$&$1.47(0.731)$&$249(32)$&$812(551)$&$2.25$&25\\
  
  Case 2 &$0.355(0.017)$&$0.021(0.003)$&$0.0(0.0)$&$0.059(0.008)$&$2.571(0.346)$&---&---&$2.08$&33\\
  Case 2, excluding 550-650d  &$0.357(0.018)$&$0.022(0.003)$&$0.0(0.0)$&$0.060(0.009)$&$2.627(0.391)$&---&---&$1.96$&24\\
 
  \hline
  \multicolumn{8}{l}{$^a$Assuming $\left(^{57}\mathrm{Fe}/^{56}\mathrm{Fe}\right)_{\sun}=0.023$ \citep{Asplund09}}\\
 
  \end{tabular}
 \end{center} 
\end{table*}

This $M(56)$ is significantly lower than other independent estimates. 
\citet{Mazzali15} estimated $M(56)=0.47\pm0.05$\,M$_{\sun}$ from
nebular spectroscopy.  \citet{Pereira13} used the bolometric light
curve around peak brightness to estimate
$M(56)=(0.44\pm0.08)\times(1.2/\alpha)$\,M$_{\sun}$, where
$\alpha=1.2\pm0.2$ is the ratio of bolometric to radioactivity
luminosities. (Instead assuming $\alpha=1.0\pm0.2$ would give
$M(56)=0.53\pm0.11$\,M$_{\sun}$.)  This suggests that the model needs
a way to lose energy outside of the optical/near-IR wavebands. We next
consider two ways in which this could occur.

\begin{figure*}
\begin{center}  
        \includegraphics[width=0.9\textwidth]{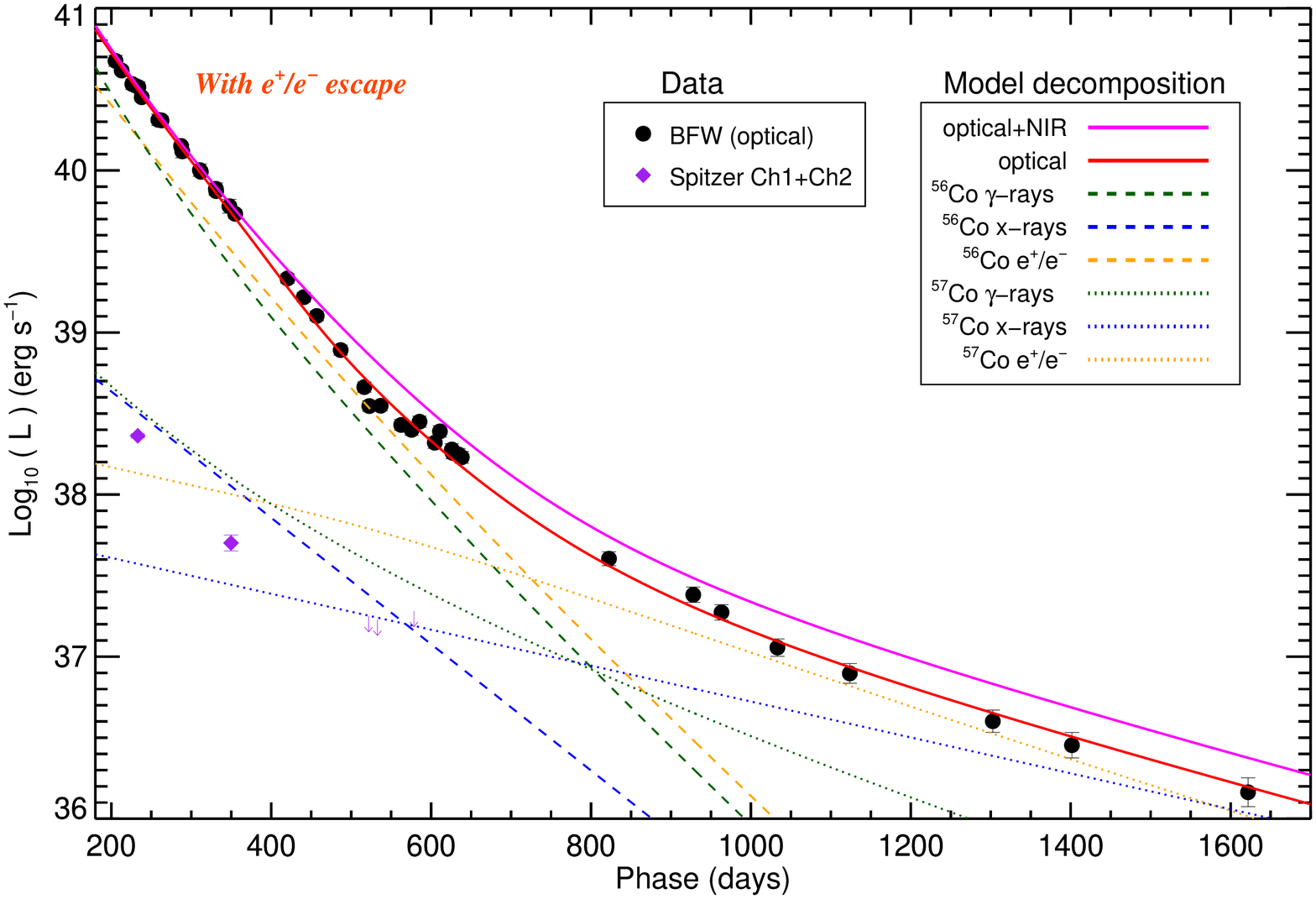}
        \caption{The pseudo-bolometric light curve of SN\,2011fe shown
          as filled black circles; see
          Section~\ref{sec:pseudo-bolomt-light} for details of the
          light curve construction. This figure shows the fit of the
          `Case 1' model (Section~\ref{sec:bol11fe}), allowing for
          positron/electron escape and assuming that the optical light curve
          and near-IR corrections (Section~\ref{sec:non-optic-contr})
          account for all of the emerging photons. 
          Fig.~\ref{fig:11fe_bolometric_light_full_fit_IRC}
          shows the fit of the `Case 2' model, which does not allow of
          positron/electron escape but instead allows for luminosity to emerge
          at wavelengths greater than 2.5\,$\micron$.}
\label{fig:11fe_bolometric_light_full_fit}
\end{center}  
\end{figure*}

\begin{figure}
\begin{center}  
        \includegraphics[width=0.5\textwidth]{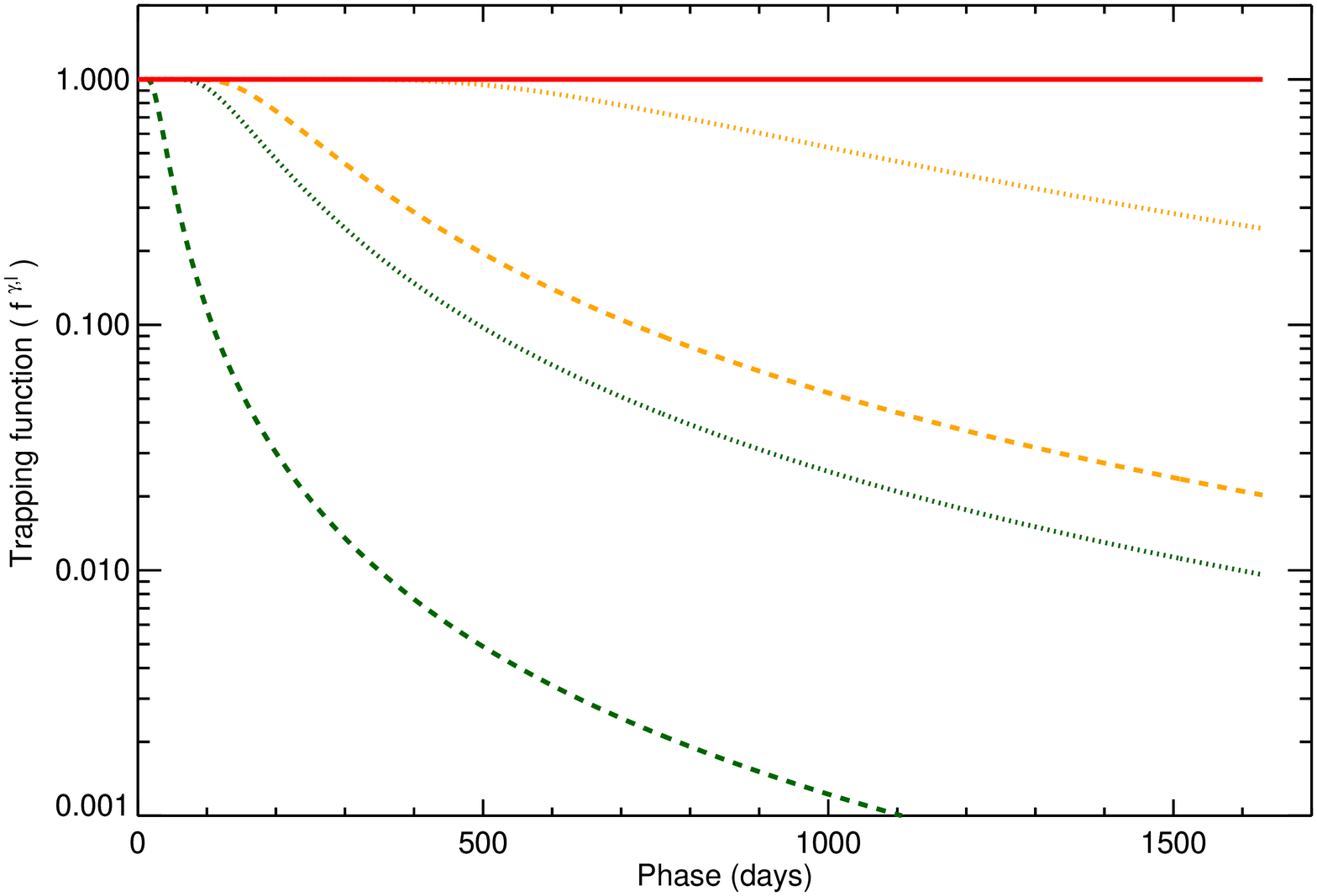}
        \caption{The inferred trapping functions for $\gamma$-rays
          ($f^{\gamma}(t)$) and positrons ($f^{l}(t)$), from
          equation~(\ref{eq:trappingfraction}), resulting from our
          'Case 1' fits. The green lines correspond to
          $\gamma$-rays with $t_{56}^{\gamma}=35$\,d (dashed) and $t_{57}^{\gamma}=160$\,d
          (dotted) that we assume for $^{56}$Co and $^{57}$Co decay respectively.  
          The yellow lines correspond to the leptonic
          trapping from the $^{56}$Co decay, $t^{l}_{56}=233$\,d (dashed), and
          the leptonic trapping for $^{57}$Co decay,
          $t^{l}_{57}=886$\,d (dotted). The red solid line corresponds to
          complete trapping.}
\label{fig:trapping_functions}
\end{center}  
\end{figure}

For \textit{Case 1}, which we refer to as the `positron/electron escape case',
we allow the positron/electron escape fractions ($t_{56}^{l}$, $t_{57}^{l}$) to
be free parameters in the fit. We then find
(Table~\ref{tab:Bol_fit_results}) $M(56)=0.461\pm0.041$\,$M_{\sun}$,
$M(57)=0.014\pm0.004$\,$M_{\sun}$ and $M(55)=0.0$\,$M_{\sun}$, with a
much improved fit quality ($\chi^2_\mathrm{DOF}\simeq2.5$ for 34 
degrees of freedom, Fig.~\ref{fig:11fe_bolometric_light_full_fit}). 
This fit provides strong evidence for the
presence of the $^{57}$Ni decay chain. The fit also requires a
considerable positron/electron escape over the phase range of 200--600\,d; for
example, at 500\,d $\simeq$75 per cent of the leptonic energetic
output escapes the ejecta. The inferred trapping functions for this
fit are shown in Fig.~\ref{fig:trapping_functions}.

Our inferred $^{56}$Ni mass, $M(56)=0.461$\,$M_{\sun}$, is consistent
with other independent estimates (see above).  The small difference
compared to that from nebular spectroscopy (around 3 per cent) could
be attributed to some part of the spectral energy distribution that is
not covered by our pseudo-bolometric filter or near-IR correction.
Generally, however, this fit provides evidence that our
pseudo-bolometric light curve and the near-IR modelling are accounting
for most of the emitted photons, \textit{assuming} that the charged leptons are
able to escape the ejecta.

We note that from around +500\,d onwards, the optical bolometric light
curve begins to decline faster, before beginning to flatten just
before +600\,d.  This behaviour, particularly the flattening, is not
well reproduced by our simple model, although the faster decline
occurs at the same time as flux is being redistributed to the near-IR
(Fig.~\ref{fig:noc_compare}). We therefore also fit the bolometric
light curve excluding the phase range of 550--650\,d, where the
underlying behaviour is not well-understood and shows a different
apparent decay rate. The parameters are again reported in
Table~\ref{tab:Bol_fit_results} -- these are close to the original fit
values, but with an improved $\chi^2_\mathrm{DOF}$.

\begin{figure*}
\begin{center}  
        \includegraphics[width=0.9\textwidth]{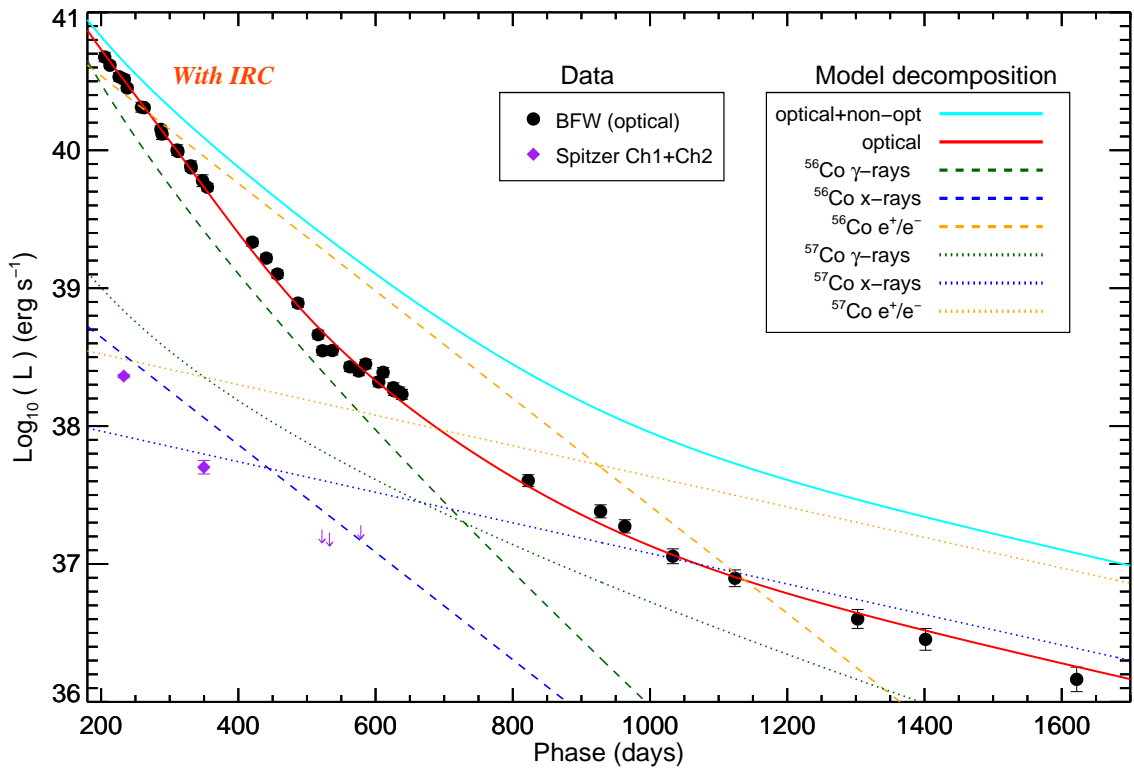}
        \caption{As Fig.~\ref{fig:11fe_bolometric_light_full_fit}, but
          for the `Case 2' model that accounts for flux emerging at
          $\lambda>2.5$\,\micron\ (i.e., an infrared catastrophe; IRC)
          and assumes full positron/electron trapping.  Here, the cyan line
          denotes the inferred total (optical+near-IR+mid/far-IR)
          bolometric light curve of the SN.}
\label{fig:11fe_bolometric_light_full_fit_IRC}
\end{center}  
\end{figure*}

Although we have accounted for the non-optical luminosity in our
fitting up to 2.5\,$\micron$, there are strong theoretical reasons to
expect significant mid/far-IR luminosity at late-times due to the IRC.
Therefore, in \textit{Case 2} we use a different approach for the
non-optical contribution.  Motivated by the fact that there has not
been any strong observational evidence for positron/electron escape in SNe Ia,
we investigate the `IRC case', where a substantial amount of flux is
shifted to far-IR wavelengths. For this, we fit a model where we force
complete positron/electron trapping for the full bolometric light curve; in
other words, we fix $f_{A}^{l}$ in equation~(\ref{eq:decaychainlum})
to 1. This final model luminosity then will be 

\begin{equation}
L_\mathrm{model}(t)=\left[L_{55}(t)+L_{56}(t)+L_{57}(t)\right]\times\left[1-F_\mathrm{non-opt}(t)\right]
\label{eq:finalmodel_IRC}
\end{equation}
where $L_{55}$, $L_{56}$ and $L_{57}$ are given as before, by
equation~(\ref{eq:decaychainlum}), and $F_\mathrm{non-opt}$ corresponds
to the \textit{total} non-optical contribution, for which we assume the same 
logistic functional form as in Fig.~\ref{fig:noc_compare}, but with a
limit at 200\,d to be $\simeq$1 per cent. We then fit for $M(56)$, $M(57)$, $M(55)$ 
and the total non-optical contribution. The result is plotted in
Fig.~\ref{fig:11fe_bolometric_light_full_fit_IRC} and again reported
in Table~\ref{tab:Bol_fit_results}.

For this fit, we find
$M(56)=0.355\pm0.017$\,$M_{\sun}$, $M(57)=0.021\pm0.003$\,$M_{\sun}$ and
$M(55)=0.0$\,$M_{\sun}$ with  $\chi^2_\mathrm{DOF}\simeq2$ for 33 
degrees of freedom. We note this fit prefers a smaller $M(56)$,
lower than previous estimates, indicating that at the earlier part of
our bolometric light curve there still is 25 per cent of the flux that
we still do not probe. Fixing instead the $M(56)$ mass to
$M(56)=0.47$\,$M_{\sun}$, and allowing the non-optical contribution at
200\,d to be unconstrained, we find $M(57)=0.028\pm0.003$\,$M_{\sun}$
and a non-optical contribution at 200\,d equal to 25 per cent, with
$\chi^2_\mathrm{DOF}\simeq2.04$ for 34 degrees of freedom.

Our inferred non-optical contribution increases to $\simeq85$ per cent
at 600\,d. This is consistent with \citet{Fransson15}, who estimated
an 80 per cent non-optical/NIR contribution at 1000\,d, although their
models indicate that this flux is mainly emitted in the mid-IR. In our
analysis, we have assumed a constant evolution of the non-optical
contributions after +600\,d, the approximate expected epoch of any
IRC. This assumption may not be correct, and a colour evolution that
decreases the near-IR flux and increases the mid/far-IR one is
certainly possible, but not constrained by our data.  We note that the
\textit{Spitzer} mid-IR flux from \citet{Johansson14} is very small,
although in the event of an IRC, the theoretical prediction is that
the fine-structure lines will be redward of the mid-IR
\textit{Spitzer} filter bandpass.

\section{Discussion}
\label{sec:discussion}

We now investigate other physical mechanisms that may contribute to
the behaviour of the bolometric late-time light curve, and then
discuss the implications of our study in the context of SN Ia
explosion models and physics.

\subsection{Other contributions to the late-time luminosity}
\label{sec:other}

Although our simple model provides a good fit to most of our
pseudo-bolometric light curve data, we cannot rule out other sources
of luminosity that may effect the late-time light curve. We discuss
these in turn \citep[see also a similar discussion in the context of
SN\,2012cg in][]{graur15}.

\subsubsection{A light echo}

The flattening of the light curve of SN\,2011fe from around 600\,d
could be explained by the presence of a dust cloud near the SN that
scatters a fraction of the SN light towards the observer, a mechanism
known as `light echo'. Due to the different light travel times, the
spectrum we observe would then be an integrated combination of the
scattered early-time spectra combined with the direct line-of-sight
late-time spectrum.  This behaviour has been demonstrated for several
SNe Ia \citep[e.g.][]{Schmidt94,Cappellaro01,2015ApJ...805...71D},
where the bolometric light curves evolve in a manner consistent with
$^{56}$Co up to $>$500--600\,d, at which point the light curves
flatten.  The spectra also evolve from classical SN Ia nebular spectra
over $\sim$300--500\,d, to spectra showing a blue continuum with broad
absorption and nebular emission features super-imposed at
$\sim$500--600\,d.  \citet{Schmidt94} and \citet{Cappellaro01} can
reproduce the late-time SN Ia spectra with a model light-echo
spectrum, constructed by co-adding early-time spectra multiplied by a
power-law scattering function, $S_{\lambda}\propto\lambda^{-\alpha}$,
with $\alpha=1,2$ \citep[for a more detailed approach
see][]{Marino15}.

\begin{figure}
\begin{center}  
        \includegraphics[width=0.5\textwidth]{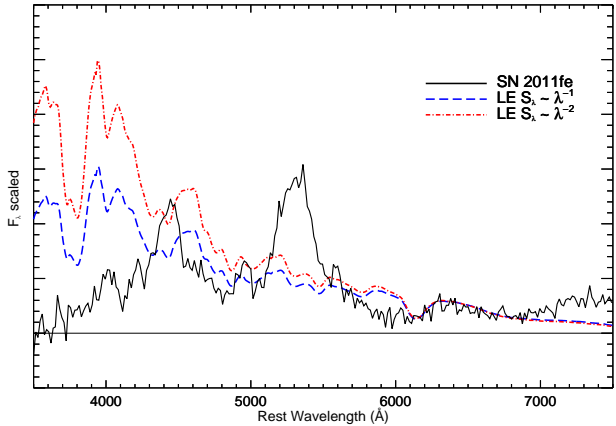}
        \caption{The $+1030$\,d SN\,2011fe spectrum \citep[black solid 
          line;][]{taubenberger15}, compared to two synthetic `light
          echo' spectra constructed with scattering laws of
          $S_{\lambda}\propto\lambda^{-1}$ (blue dashed line) and
          $S_{\lambda}\propto\lambda^{-2}$ (red dashed-dotted line). For details see
          Section~\ref{sec:other}.}
\label{fig:11fe_le}
\end{center}  
\end{figure}

While the late-time photometric behaviour of SN\,2011fe partly
resembles a light echo (e.g., the flattening at $\sim$600\,d), a light
echo is unlikely to be responsible. The late-time spectra are
considerably different to a blue-continuum light echo spectra. We
demonstrate this following the procedure of \citet{Schmidt94} and
\citet{Cappellaro01}.  We used the available early-time spectra of
SN\,2011fe from \textit{WISEREP}, spanning $-5$ to $+75$\,d, and after
weighting them by the integrated luminosity at their observed epochs,
correcting for reddening and using scattering laws as above, we co-add
them constructing an expected light echo spectrum. In
Figure~\ref{fig:11fe_le}, we compare the late-time SN\,2011fe spectrum
with the two synthesised light echo spectra for the two different
scattering laws with $\alpha=1,2$. We observe no similarities between
the observed and synthetic spectra, and thus conclude that a light
echo cannot significantly influence the SN\,2011fe light curve.
A similar study has been performed by \citet{Shappee16},
  reaching similar conclusions.

\subsubsection{Circumstellar material}

A possible explanation for the brightness of SN\,2011fe at these late epochs
is that the light curve is contaminated by interaction of the ejecta
with circumstellar material (CSM) around the explosion. While a
complete investigation of these possibilities is beyond the scope of
this paper, we note that we do not see hydrogen in any of the
late-time spectra, indicating little interaction with a CSM, with no
hot black-body component and no narrow/intermediate emission lines
of any element.

\subsubsection{A surviving companion}

Finally, we consider whether a surviving binary companion could also
contribute significant luminosity at late times. If we fit the
bolometric light curve with a simple $^{56}$Co decline law and a
constant luminosity component, simulating a surviving companion, we
find $L_\mathrm{companion}=7.7\pm0.6\times10^{2}$\,$L_{\sun}$, with
$\chi^2_\mathrm{DOF}\simeq8.6$ for 36 degrees of freedom.  This upper
limit would be consistent with a red giant or main-sequence star of
$\sim6$\,M$_{\sun}$, but such a possibility is excluded by the
analysis of the \textit{HST} pre-explosion imaging by \citet{Li11} and
stellar evolution timescales (however, less massive companions could
have an increased luminosity due to interaction with the SN ejecta).
Thus, again, this possibility seems unlikely, again as was
  also found in \citet{Shappee16}.

\subsection{Implications}
\label{sec:implications}

\subsubsection{The infrared catastrophe or positron/electron escape?}
\label{sec:irc}

We next investigate whether an infrared catastrophe (IRC) may have
occurred in SN\,2011fe, and discuss the implications of this
phenomenon regarding the late light curve evolution. As discussed in
the introduction, the IRC is the result of a rapid cooling in the
ejecta, with the thermal emission moving from the optical/near-IR to
the mid/far-IR.  There is no published direct evidence for a dramatic IRC in
SN\,2011fe, either in the spectroscopy, where prominent optical
emission lines are still seen \citep[e.g.,][]{taubenberger15}, or in
other photometric studies \citep{Kerzendorf14}.

However, in this paper, we have shown that the pseudo-bolometric light curve is difficult to reconcile with the
evolution expected from the $^{56}$Ni and $^{57}$Ni decay chains with models involving complete or nearly-complete
trapping of the charged leptons. Such a model can be made to be
consistent with the data in the presence of significant positron/electron
escape.

We note that such a positron escape is disfavoured, since it would be
difficult to explain the distribution of the observed positron signal from the Milky Way. \citet{Crocker16} 
determine the required age of a putative 
stellar source of the Galactic positrons, constrained by the recently revised 
positron annihilation signal of the bulge, disc, and nucleus of our Milky Way 
\citep{Siegert16}. \citet{Crocker16} require the putative stellar source of 
the positrons to have an age of ~3-6 Gyr and argue strongly against normal 
SNe Ia (with typical ages of 300 - 1000 Myr) as the main source of positrons 
in our Galaxy.

Therefore, taking into account the lack of any evidence for very high positron/electron 
escape from SNe Ia, we favour the interpretation of the late time light curve evolution
as a redistribution of flux from the optical/NIR to the mid/far-IR, similar to an occurrence
of an IRC.

\subsubsection{Evidence for $\rm{^{57}Ni}$ and $\rm{^{55}Co}$}
\label{sec:ni57}

The luminosity produced by the $^{57}$Ni decay chain is unambiguously
detected in our analysis; fits based just on $^{56}$Ni cannot
reproduce the observed evolution. In principle, a measurement of the
ratio of the synthesised $^{57}$Ni to $^{56}$Ni masses
($M(57)$/$M(56)$) can possibly disentangle potential explosion
mechanisms, and thus provide some insight into the progenitor system.
A recent demonstration on the SN Ia SN\,2012cg \citep{graur15} claimed
$M(57)/M(56)=1.87^{+0.52}_{-0.48}$ \citep[here we give all ratios in
terms of the solar value of
$\left(^{57}\mathrm{Fe}/^{56}\mathrm{Fe}\right)_{\sun}=0.023$; see][]{Asplund09},
increasing to a higher value of $2.96^{+0.83}_{-0.78}$ when enforcing
$M(55)=0$.

However, such measurements are clearly susceptible to assumptions made
in the modelling, and thus systematic uncertainties. Our fit ratios
range from $M(57)/M(56)=1.3$--2.5, i.e. super-solar.  These values are
broadly consistent with the ratios predicted for the one-dimensional
W7 \citep[1.7;][]{Iwamoto99}, the two-dimensional
\citep[1.4-1.9;][]{Maeda10} and three-dimensional
\citep[1.3;][]{Seitenzahl13} delayed detonation models, but not with
the violent merger \citep[1.1;][]{Pakmor12} model. 

While the systematic uncertainties of our modelling are large, the (simple)
models that we have investigated indicate a high $M(57)/M(56)$ ratio, in contradiction with \citet{Shappee16},
pointing to an explosion mechanism with relatively high central
density of synthesised $\rm{^{56}Ni}$. This kind of explosion
mechanisms are associated with SD progenitor channels (however, see
\citealt{Fenn16}).

In all of our fits, we use a physical prior in our models that all of
the synthesised masses must be greater than or equal to zero. The fits
including $^{55}$Co hit this limit for $M(55)$, precluding the
calculation of uncertainties in this parameter. Relaxing this
assumption gives a very small negative value for $M(55)$ with an
uncertainty of $6.9\times10^{-3}$\,M$_{\sun}$ and
$1.6\times10^{-2}$\,M$_{\sun}$ for `Case 1' and `Case 2',
respectively. The corresponding upper limit ratios are then
$M(55)/M(57)=0.47$ and $0.56$; the corresponding predicted ratios for
the two-dimensional delayed detonation models are $0.4-0.6$
\citep{Maeda10}, for the three-dimensional delayed detonation model
\citep{Seitenzahl13} the prediction is $0.7$, and for the violent
merger model \citep{Pakmor12,ropke12} $0.25$.


\section{Conclusions}
\label{sec:conclusions}

In this paper we have presented a late-time optical $R$-band light
curve for the type Ia supernova SN\,2011fe, measured using data from
the Palomar Transient Factory. Combining this light curve with other
published photometric data from ground-based telescopes and the
\textit{Hubble Space Telescope}, and with ground-based spectroscopy,
we have estimated the pseudo (`optical') bolometric light curve out to
1600 days after peak brightness. We also construct a model for the
likely bolometric output in the near-IR. We analyse this light curve
using a simple model of energy deposition from three possible decay
chains ($^{56}$Ni, $^{57}$Ni and $^{55}$Co). Our main findings are:

\begin{itemize} 

\item The bolometric light curve for SN\,2011fe can be broadly
  explained by the radioactive inputs from the $^{56}$Ni and $^{57}$Ni
  decay chains. The presence of the $^{57}$Ni decay chain is required
  and is robustly detected in our analysis. 
\item The PTF $R$-band light curve has a noticeable rapid decline
  followed by a plateau over 500-600 days. While the rapid decline can
  be explained by a cooling ejecta and the emergence of [\ion{Fe}{ii}]
  lines producing increased near-IR flux, our simple models are unable
  to adequately explain the plateau feature in the light curve.
\item Our pseudo-bolometric light curve is not consistent with models
  that have full trapping of the produced charged leptons and no infrared
  catastrophe (IRC). An additional route for energy escape outside of
  the optical/near-IR is required.
\item The bolometric light curve is consistent with models that allow
  for positron/electron escape. In this case, our best-fitting initial
  $^{56}$Ni mass is $M(55)=0.461\pm0.041$\,$M_{\sun}$, consistent with
  independent estimates from nebular spectroscopy analysis and
  Arnett's Law. However this model requires 75 per cent of the
  charged leptons to escape by 500\,d.
\item Alternatively, our data are also fully consistent with a model
  that has complete positron/electron trapping but does allow for a redistribution
  of flux to the mid-far IR, similar to an IRC \citep{Axelrod80}. In this
  case, around 85 per cent of the total bolometric luminosity must be
  escaping at non-optical wavelengths by day 600 and onwards.
\item For both of these scenarios, the amount of $^{57}$Ni we estimate
  is relatively large, compared to popular explosion models. This
  large inferred mass indicates high central density explosion
  environments, mainly predicted from SD progenitor channels and WD collisions.
\item Including contributions from the $^{55}$Co decay chain does not
  improve the quality of the bolometric light curve fits, although
  this is not well constrained by the dataset. We estimate an upper
  limit of $^{55}$Co mass of $1.6\times10^{-2}$\,M$_{\sun}$.
\end{itemize}

This paper has highlighted the significant systematic uncertainties
that are involved in modelling the late-time light curves of SNe Ia,
even within a very simple framework. Even though high-quality data
were obtained for SN\,2011fe during the late phases, a denser time
sampling of photometric and spectroscopic late-time observations
spanning the optical and infrared are needed for future similar
events, particularly over the phase range of 400--800 days, where the
ionisation state appears to change and a sharp colour evolution can
occur (including mid and far-IR data where possible). We also
neglected dependent effects in our modelling, such as freeze-out.
While these effects are predicted to be small out to at least 900 days
\citep{Fransson15}, more detailed modelling may provide an improved
interpretation of the very late-time light curve. Repeating this study
for a larger sample of events will allow us to precisely measure and
compare the nickel masses obtained from peak and late times,
extracting information on the energetic output of the radioactive
decay and confirm whether positron escape and/or an IRC is occurring.

\section*{Acknowledgments}

The authors thank Claes Fransson and Anders Jerkstrand for helpful
discussions regarding the infrared catastrophe and its theoretical
implications.

We acknowledge support from EU/FP7-ERC grant No. [615929] and the
Weizmann-UK `Making Connections' program. This research was supported
by the Munich Institute for Astro- and Particle Physics (MIAPP) of the
DFG cluster of excellence ``Origin and Structure of the Universe''.
Part of this research was carried out at the Jet Propulsion
Laboratory, California Institute of Technology, under a contract with
the National Aeronautics and Space Administration.

Based on observations made with ESO Telescopes at the La Silla Paranal
Observatory under programme IDs 091.D-0764, 092.D-0632 and 096.D-0627.
Based on observations made with the NASA/ESA \textit{Hubble Space
  Telescope}, obtained from the Data Archive at the Space Telescope
Science Institute, which is operated by the Association of
Universities for Research in Astronomy, Inc., under NASA contract NAS
5-26555. These observations are associated with programs \#13737 and
\#14166. This research has made use of the NASA/IPAC Infrared Science
Archive, which is operated by the Jet Propulsion Laboratory,
California Institute of Technology, under contract with the National
Aeronautics and Space Administration.


\bibliographystyle{mnras}
\bibliography{11fe_late_time_light_curve}

\appendix
\section{Photometric data}
\label{sec:phot_spec_data}

\begin{table*}{}
\caption{PTF Photometry for SN\,2011fe.}
\label{tab:11fe_phot_log_ptf}
\begin{tabular}{@{}lcccccl}
\hline
   MJD&Phase (days)$^{a}$&Telescope&Filter&Mag(AB)&$\rm{\Delta}$Mag(AB)\\
\hline
$56016.138$&$201.59$&P48&PTF48R&$16.034$&$0.009$\\
$56016.201$&$201.65$&P48&PTF48R&$16.045$&$0.016$\\
$56016.263$&$201.72$&P48&PTF48R&$16.058$&$0.008$\\
$56017.170$&$202.62$&P48&PTF48R&$16.085$&$0.009$\\
$56017.214$&$202.67$&P48&PTF48R&$16.077$&$0.009$\\
$56017.245$&$202.70$&P48&PTF48R&$16.073$&$0.008$\\
$56017.463$&$202.91$&P48&PTF48R&$16.050$&$0.008$\\
$56017.507$&$202.96$&P48&PTF48R&$16.064$&$0.008$\\
...&...&...&...&...&...\\
...&...&...&...&...&...\\
...&...&...&...&...&...\\
 \hline
 \multicolumn{6}{l}{$^a$MJD$_\mathrm{max}=55814.38$, calculated by SiFTO}\\
 \end{tabular}
\end{table*}

\bsp
\label{lastpage}
\end{document}